\documentclass{aa}  
\usepackage{graphicx}
\usepackage{caption}
\usepackage{subcaption}

\usepackage{txfonts}
\usepackage{xcolor}
\usepackage[colorlinks = true,
            linkcolor = blue,
            urlcolor  = blue,
            citecolor = blue,
            anchorcolor = blue,
            breaklinks=true]{hyperref}

\makeatletter
\renewcommand*\aa@pageof{, page \thepage{} of \pageref*{LastPage}}
\makeatother

\usepackage{blindtext}

\usepackage{booktabs}
\usepackage{placeins}

\usepackage{multirow}
\usepackage{amsmath}
\usepackage{lscape}

\def\degr{\hbox{$^\circ$}}

\def\arcsec{\hbox{$^{\prime\prime}$}}

\def\rfive{\hbox{$R_\mathrm{500}$}}

\def\mujybeam{\hbox{$\mu\mathrm{Jy\,beam}^{-1}$}}

\usepackage[nolist]{acronym}

\newacro{RM}[RM]{rotation measure}
\newacro{AGN}[AGN]{active Galactic nuclei}
\newacro{VLA}[VLA]{Karl G. Jansky Very Large Array}
\newacro{ICM}[ICM]{intracluster medium}
\newacro{PSZ2}[PSZ2]{Planck 2nd Sunyaev-Zeldovich Source Catalog}
\newacro{MCMC}[MCMC]{Monte Carlo Markov chain}
\newacro{CC}[CC]{cool-core}
\newacro{NCC}[NCC]{non-cool-core}

\begin{document}

   \title{Probing cluster magnetism with embedded and background radio sources in Planck clusters}


\author{E. Osinga
        \inst{1,2}
        \and 
        R. J. van Weeren\inst{2}
        \and 
        L. Rudnick\inst{3}
        \and
        F. Andrade-Santos\inst{4,5,6}
        \and
        A. Bonafede\inst{7,8}
        \and
        T. Clarke\inst{9}
        \and
        K. Duncan\inst{10}
        \and
        S. Giacintucci\inst{9}
        \and
        H. J. A. R\"ottgering\inst{2}
        }

\institute{Dunlap Institute for Astronomy \& Astrophysics, University of Toronto, 50 St. George Street, Toronto, ON M5S 3H4, Canada \\ \email{erik.osinga@utoronto.ca}
\and 
Leiden Observatory, Leiden University, PO Box 9513, NL-2300 RA Leiden, The Netherlands 
\and 
Minnesota Institute for Astrophysics, University of Minnesota, 116 Church St SE, Minneapolis, MN 55455, USA 
\and 
Center for Astrophysics | Harvard \& Smithsonian, 60 Garden Street, Cambridge, MA 02138, USA 
\and 
Department of Liberal Arts and Sciences, Berklee College of Music, 7 Haviland Street, Boston, MA 02215, USA 
\and 
Clay Center Observatory, Dexter Southfield, 20 Newton Street, Brookline, MA 02445, USA 
\and 
DIFA - Universit\`a di Bologna, via Gobetti 93/2, I-40129 Bologna, Italy 
\and
INAF - IRA, Via Gobetti 101, I-40129 Bologna, Italy; IRA - INAF, via P. Gobetti 101, I-40129 Bologna, Italy 
\and 
Naval Research Laboratory, 4555 Overlook Avenue SW, Code 7213, Washington, DC 20375, USA 
\and 
Institute for Astronomy, Royal Observatory, Blackford Hill, Edinburgh, EH9 3HJ, UK 
}

   \date{Received 2024-08-13; accepted Y}

 
  \abstract
    {Magnetic fields remain an elusive part of the content of galaxy clusters. Faraday rotation and depolarisation of extragalactic radio sources are useful probes, but the limited availability of polarised radio sources necessitates the stacking of clusters to study average magnetic field properties. We recently presented a Karl G. Jansky Very Large Array survey of the 124 most massive Planck clusters at low redshift ($z<0.35$), finding a clear depolarisation trend with the cluster impact parameter, with sources at smaller projected distances to the cluster centre showing more depolarisation. In this study, we combine the depolarisation information with the observed rotation measure (RM) and present an investigation into the average magnetic field properties of the sample, using both background sources and sources embedded in clusters. We observe a significant increase in the RM scatter, $\sigma_\mathrm{RRM}$, closer to the cluster centres. Averaging all 124 clusters, we find a scatter within $R_\mathrm{500}$ of $\sigma_\mathrm{RRM}=209\pm37$ rad m$^{-2}$, with background sources and cluster members showing similar values ($200\pm33$ and $219\pm66$ rad m$^{-2}$, respectively). In the simple assumption of a uniform amplitude magnetic field with a single fluctuation scale $\Lambda_c$, this translates to an average magnetic field strength of $2\,(\Lambda_c/10\mathrm{kpc})^{-0.5}\, \mu$G. The profile of $\sigma_\mathrm{RRM}$ as a function of projected radius is inconsistent with a model that has a simple scaling $B \propto n_e^\eta$, with an observed deficit near the centre of clusters possibly caused by the fact that the highest RM sources near the centre of clusters are depolarised. Combining depolarisation and RM in a full forward model, we find that the magnetic field power spectrum roughly agrees with the Kolmogorov value, but that none of the Gaussian random field models can fully explain the observed relatively flat profiles. This implies that more sophisticated models of cluster magnetic fields in a cosmological context are needed.
    }

   \keywords{magnetic fields – polarization – galaxies: clusters: general – galaxies: clusters: intracluster medium – radiation mechanisms: non-thermal – methods: observational}

   \maketitle
%


\section{Introduction}\label{sec:introduction}
Galaxy clusters, the largest gravitationally bound structures in the Universe, harbour a rich variety of physical phenomena. Radio observations have revealed that clusters often show diffuse synchrotron emission that can span Mpc-sized regions, such as `radio halos' \citep[e.g.][]{Bonafede2022} or `mega-halos' \citep[][]{Cuciti2022}, implying that clusters are filled with ultra-relativistic electrons and magnetic fields. The influence of the magnetic fields extends to particle acceleration models, radio synchrotron age estimates, the dynamics of the \ac{ICM} and the transport of cosmic rays. Understanding the properties and origins of magnetic fields in clusters thus has broad importance \citep[see][for reviews on magnetic fields in galaxy clusters]{Carilli2002, GovoniFeretti2004, Donnert2018}. 

The most promising tool to study magnetic fields is radio polarisation observations. A magnetised plasma such as the \ac{ICM} causes a wavelength-dependent rotation of the polarisation angle (Faraday rotation) and depolarisation. In general, the Faraday depth of a source is defined as \citep{Burn1966,Brentjens2005}
\begin{equation}
    \phi(\textbf{r}) = 812 \int_\mathrm{LOS} n_e \textbf{B} \cdot d\textbf{r} \; \mathrm{rad\: m}^{-2},
\end{equation}
where $n_e$ is the electron density in parts per cm$^{-3}$, $\textbf{B}$ is the magnetic field in $\mu$Gauss and d$\textbf{r}$ the infinitesimal path length increment along the line of sight (LOS) in kpc, and we define $\phi(\textbf{r})>0$ for the magnetic field pointing towards the observer. In the simple case of just one radio-emitting source along the line of sight, the Faraday depth is equal to the \ac{RM}. With a combination of {RMs from radio observations and electron densities from X-ray observations}, it is thus possible to study the magnetic field properties of galaxy clusters. 

Such studies are best done at low redshifts to maximise the cluster angular size because polarised radio sources are relatively rare \citep[e.g.][]{Rudnick2014}. The most detailed analyses have been of the Coma Cluster \citep{Bonafede2010} and Abell 2345 \citep{Stuardi2021}, where seven polarised radio sources were detected per cluster. The Coma Cluster magnetic field was found to agree with a Kolmogorov power spectrum with a central strength of 5 $\mu$G and a scaling of magnetic field energy density linearly proportional to the thermal gas density ($B^2 \propto n_e$). The central magnetic field strength in Abell 2345 was found to be similar to the Coma Cluster, but with a magnetic field energy density that scales {more steeply} instead (i.e. $B^2 \propto n_e^2$). 
Several other low-redshift clusters have been analysed in polarisation \citep{Murgia2004,Govoni2006,Guidetti2008,Govoni2010,Vacca2012,Govoni2017}, with typically less than five polarised radio galaxies per study, resulting in large uncertainties on the magnetic field estimates \citep[see e.g.][for a detailed discussion]{Johnson2020}.

Because cluster magnetic fields are thought to be generally turbulent and disordered, the observed Faraday rotation is the result of a random walk process and thus a random variable. Accurate magnetic field estimates, therefore, require a statistical analysis probing many independent sight lines. Another potential problem is that polarised radio galaxies are often embedded in the cluster, and the degree to which the observed RM variations are caused by local interaction of the lobes with the ICM is debated \citep{Rudnick2003,Laing2008,Guidetti2012,Osinga2022}. Such problems can be overcome by stacking clusters to increase the number of polarised radio sources located behind clusters and thus the number of independent sight lines through a cluster \citep{Clarke2001,Bonafede2011,Bohringer2016,Stasyszyn2019,Osinga2022}. Although stacking experiments have limited ability to probe differences between clusters, they are useful for obtaining average cluster magnetic field properties and are currently the only way to study cluster magnetic fields beyond the few nearest clusters. 

We recently published the largest homogeneous stacking experiment using \ac{VLA} observations of 124 galaxy clusters \citep{Osinga2022} selected from the \ac{PSZ2} \citep{Planck2016}. This study presented the first clear depolarisation trend tracing the radial profile of cluster magnetic fields using over 600 polarised radio sources. While depolarisation traces the smaller scale structure of the magnetic fields {on the scale of the resolving beam} (usually kpc), the larger scale structure {(i.e. along the line of sight, usually Mpc)} can be inferred from the Faraday rotation of the radio sources. In this paper, we add the information from the Faraday rotation of the same sample of sources to study the large-scale properties of the magnetic fields in galaxy clusters. By jointly fitting both depolarisation and Faraday rotation, we aim to constrain the average magnetic field strength, scaling with density, and power spectrum. Cosmological calculations are performed assuming a flat $\Lambda$CDM model with $H_0=70$ kms$^{-1}$Mpc$^{-1}$, $\Omega_m=0.3$ and $\Omega_\Lambda=0.7$ and the $R_\mathrm{500}$ radius is the radius enclosing an overdensity of $500$ at the cluster redshift.

\section{Chandra-Planck ESZ sample}\label{sec:data}
The sample of galaxy clusters is a subset of 124 out of 165 clusters from the \textit{Chandra-Planck Legacy Program for Massive Clusters of Galaxies}\footnote{\url{http://hea-www.cfa.harvard.edu/CHANDRA_PLANCK_CLUSTERS/}} \citep{AndradeSantos2021} that have \ac{VLA} observations presented by \citet{Osinga2022}. The full details on the data reduction, polarised source identification and association, and determination of the polarisation properties are presented in the aforementioned paper, but we briefly summarise the important points here and highlight some improvements to the catalogue. All clusters are at low redshift ($z=0-0.35$) to maximise their apparent size. The analysis of the Chandra data and resulting electron density profiles is presented in \citet{AndradeSantos2017}.

Each of the 124 clusters was observed for $\sim 40$ min in the \ac{VLA} L-band (1--2 GHz), resulting in typically 20--30 \mujybeam\,noise levels at a resolution of 6--7\arcsec\, after data reduction. Polarised sources were identified using RM-synthesis \citep{Brentjens2005} and matched to total intensity components and optical counterparts. Optical counterparts were used to determine cluster membership, with spectroscopic redshifts where available and photometric redshifts otherwise. {The full details on redshift determination are given in \citet{Osinga2022}, where Eq. 25-26 were used to divide sources into background and cluster members.}
In total 6{,}807 and 819 source components were detected in total and polarised intensity respectively. Throughout this paper we will simply refer to these components as `sources'. We have fit the following model to the polarised intensity as a function of wavelength $\lambda$, which accounts for rotation and depolarisation
\begin{equation}\label{eq:PvsLamda2}
    P(\lambda^2) = p_0 I\exp (-2\Sigma^2_{\mathrm{RM}}\lambda^4)\exp[2i(\chi_0+\phi \lambda^2)],
\end{equation}
where $p_0$ denotes the intrinsic polarisation and $\Sigma^2_{\mathrm{RM}}$ the inferred variance of the \ac{RM} distribution within the observing beam which models the depolarisation as a function of wavelength\footnote{We note that \citet{Osinga2022} defined this as $\sigma^2_\mathrm{RM}$, but we use a different nomenclature here to avoid confusion with the scatter in RM between sources.}. $\chi_0$ is the intrinsic polarisation angle, and $I$ denotes the total intensity model, which was assumed to be a simple power-law of the form $I(\nu) = I_0 \nu^\alpha$ where $\nu$ is the frequency and $\alpha$ the spectral index. {Equation \ref{eq:PvsLamda2} assumes that \citep[see][for details]{1998MNRAS.299..189S}:}
\begin{enumerate}
    \item {The dominant mechanism causing Faraday rotation is an external Faraday screen (i.e. a non-emitting foreground magnetised plasma, such as the ICM). }
    \item {The foreground screen magnetic field changes on scales (at least) smaller than the observing beam, which causes Faraday dispersion inside the beam, resulting in depolarisation.}
    \item {The magnetic field fluctuates as a Gaussian random field, implying its scatter can be completely captured in the parameter $\Sigma_\mathrm{RM}$. }
\end{enumerate}
{It has been shown \citep{Osinga2022} that this model accurately fits 75\% of the polarised radio sources detected in this sample, and we discard sources that do not show a proper fit (see below).}

We have improved the \ac{MCMC} fitter used by \citet{Osinga2022} to now properly take into account the circular nature of $\chi_0$ during the fitting. \citet{Osinga2022} used a uniform prior on $\chi_0$,  $\sim \mathcal{U} (0,\pi)$, which would cause the sampler in some cases to become trapped around the boundary values. We removed this prior on $\chi_0$ and fold the chain back into the range $[0,\pi)$ after the sampling is completed. We also calculate the mean and spread using circular statistics where we take into account the fact that the angles are distributed on the half-circle $\in$ $[0,\pi)$. 
In this way, the mean denotes the angle of the average vector on the unit circle, and the standard deviation is the spread in angles around the average vector. This agrees with the definition of the simple arithmetic mean and standard deviation when the angles are distributed away from the edges of the domain \citep[{see e.g.}][{for more details on circular statistics}]{mardia2009directional}. 

\begin{table}[hbt]
\caption{Number counts of polarised sources and unique clusters in which they were detected for different subsets of the catalogue in Appendix \ref{app:polcatalogue}.}
\label{tab:polsourcesnums}
\resizebox{\columnwidth}{!}{%
\begin{tabular}{@{}lll@{}}
\toprule
Subset                  & Number of polarised sources & Number of clusters \\ \midrule
All                     & 819                         & 124                \\
Used in analysis        & 610                         & 117                \\
Sources inside clusters & 231                         & 67                 \\
Sources behind clusters & 363                         & 101                \\
$r/R_\mathrm{500}<1$    & 261                         & 68                 \\
$r/R_\mathrm{500}<1$ with X-ray    & 233                         & 56                 \\ \bottomrule
\end{tabular}%
}
\tablefoot{Only clusters with X-ray data were forward-modelled (Sec. \ref{sec:RMratio} \& \ref{sec:forwarmodel}).}
\end{table}

This paper is accompanied by an updated table of polarised components, shown in Appendix \ref{app:polcatalogue}. We note that this update mainly corrects the quoted mean and uncertainty of the intrinsic polarisation angle $\chi_0$ and most sources have {similar best-fit RM ($\phi$) and depolarisation ($\Sigma_\mathrm{RM}$)} parameters {to within a few per cent}. This thus does not significantly impact the results. Finally, we used the same criteria for identifying bad fits as \citet{Osinga2022}. All sources with a best-fit $\chi^2$ value that is $>5\sigma$ away from the theoretical distribution, and sources with low signal-to-noise polarised emission resulting in artificially large values of $\Sigma_\mathrm{RM}^2$ were flagged. This led to 195 bad fits among the 819 polarised sources identified in 124 galaxy clusters. After excluding three clusters (G121.11+57.01, G115.71+17.52, G139.59+24.18) due to a lack of accurate redshift estimates, a total of 610 sources across 117 clusters were available for subsequent analysis. Table \ref{tab:polsourcesnums} provides an overview of the number counts for various sub-samples and the unique clusters in which these sources were found, aiding in understanding the selection function of the defined subsets in the following analysis.

\section{Methods}\label{sec:methods}
\begin{figure*}[tbh]
\centering
\includegraphics[width=0.49\textwidth]{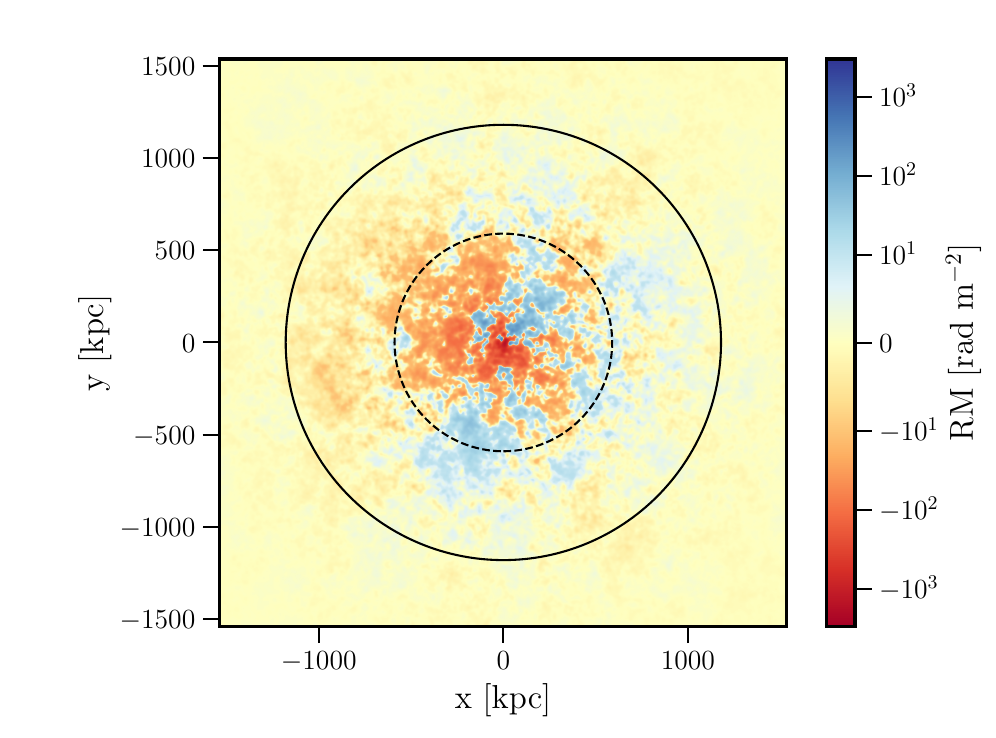} 
\includegraphics[width=0.49\textwidth]{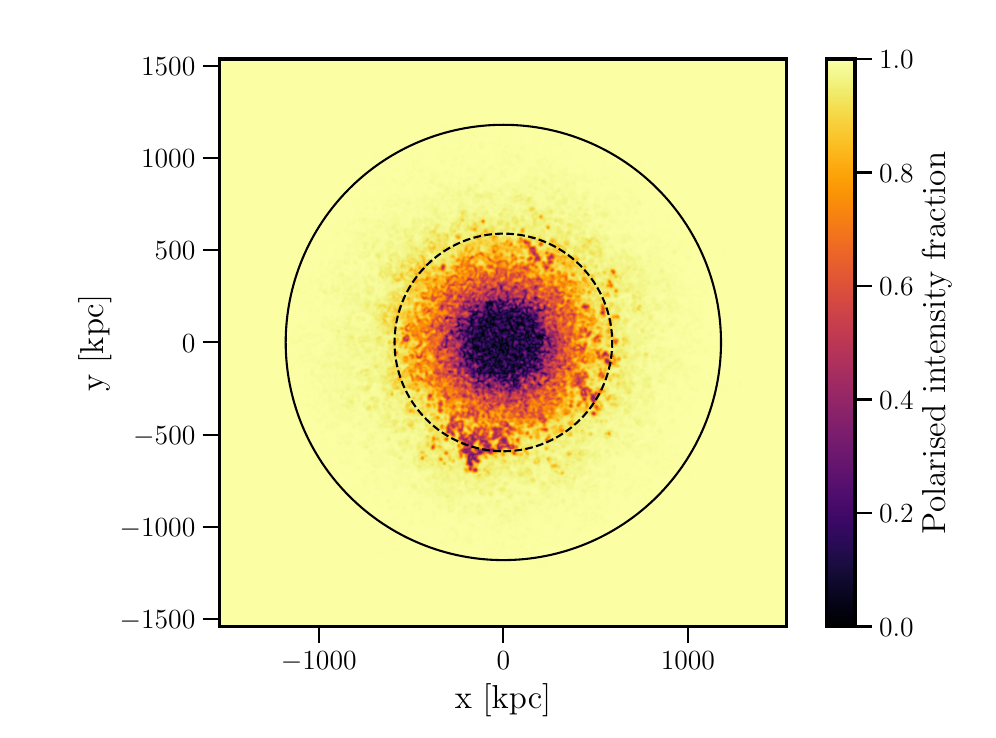}
\caption{Simulated Faraday rotation (left) and depolarisation (right) screen for a random cluster from our sample as it would appear at 1.5 GHz with perfect sampling. The parameters are $B_0=5.0\, \mu$G, $n=2$, $\eta=0.5$ and the dashed and solid circles show $0.5\rfive$ and $1.0\rfive$, respectively. \vspace{-2mm}}
\label{fig:simulation}
\end{figure*}

\subsection{Determining RM scatter}
We first provide a brief note on nomenclature to ensure clarity regarding the terminology associated with various types of scatter. We adopt the following definitions throughout this paper: 

\begin{itemize}
    \item $\delta$RM for the measurement uncertainty on RM.
    \item $\sigma_\mathrm{RM}$ for the measurable scatter (i.e. standard deviation) in RM between sources.
    \item $\Sigma_\mathrm{RM}$ for the inferred RM fluctuation across sources associated with beam depolarisation.
    \item $\delta\sigma_\mathrm{RM}$ for uncertainty on the scatter in RM ($\sigma_\mathrm{RM}$).
    \item $\delta\Sigma_\mathrm{RM}$ for uncertainty on $\Sigma_\mathrm{RM}$.
\end{itemize}
The observed (\textbf{noisy)} RM of a source, RM$_\mathrm{obs}'$, not only encompasses RM$_\mathrm{cluster}$ induced by the ICM but also contains the intrinsic RM of the source ($\mathrm{RM}_\mathrm{src}$), the RM picked up as the light travels through the intergalactic medium (IGM), the RM introduced by the Galactic foreground (RM$_\mathrm{gal}$) and finally the Earth's ionosphere (RM$_\mathrm{ion}$). Taking into account cosmological redshift, we can define
\begin{align}
\begin{split}
    \mathrm{RM}_\mathrm{obs'} &= \frac{\mathrm{RM}_\mathrm{src}}{(1+z_\mathrm{intr})^2} + \mathrm{RM}_\mathrm{IGM} + \frac{\mathrm{RM}_\mathrm{cluster}}{(1+z_\mathrm{cluster})^2} \\
    & \quad + \mathrm{RM}_\mathrm{gal} + \mathrm{RM}_\mathrm{ion},
\end{split}
\end{align}
{where $\mathrm{RM}_\mathrm{IGM}$ is the integrated contribution of the IGM, which itself has a redshift dependence \citep[c.f.][Eq. 2]{Amaral2021} that is however not relevant for this work as we regard it as a single nuisance parameter.}


The scatter in RM between sources, or RM variance, analogously is the sum of the variance of these individual terms, as they are independent contributions. The measurement noise is known from the estimated uncertainties in RM from the MCMC spectral fitting \citep[{we use the $16$th and $84$th percentile of the posterior, see}][{for details}]{Osinga2022}. The Galactic foreground can be largely removed by correcting the RMs for the Galactic contribution, which has been mapped most recently by \citet[][]{Hutschenreuter2022}. The most difficult components to isolate are the variance caused by the IGM and the intrinsic RM variance of radio sources. Studies using low RM variance sources (i.e. away from clusters and groups) that attempt to isolate the extragalactic from the Galactic and intrinsic scatter find that the extragalactic RM (rms) scatter is between $\sim$ 1 and 10 rad m$^{-2}$ \citep{Oppermann2015,Vernstrom2019,OSullivan2020,Pomakov2022}.


In this study, we aim to separate the cluster RM scatter from the other sources of scatter. The cluster RM is often the largest term, as near the centre of clusters, RMs can be on the order of $10^3$ rad m$^{-2}$. Meanwhile, the Galactic contribution to the RM is expected to be small since the cluster sample is selected from the \ac{PSZ2} survey and thus avoids the Galactic plane by design (Galactic latitude $\vert b\vert > 14\degr$). We subtracted the Galactic contribution (on the order of $10^1$ rad m$^{-2}$) using the recent map from \citet{Hutschenreuter2022}, propagating the uncertainties in quadrature.  


After subtraction of the Galactic contribution, we can define the residual RM as $\mathrm{RRM} = \mathrm{RM}_\mathrm{obs'}-\mathrm{RM}_\mathrm{gal}$, and are thus left \textbf{with}
\begin{equation}\label{eq:RRM}
    \mathrm{RRM} = \frac{\mathrm{RM}_\mathrm{cluster}}{(1+z_\mathrm{cluster})^2} + \mathrm{RM}_\mathrm{extr}
\end{equation}
where we defined the last term as the extrinsic RM, which captures the RM contribution that is not from the cluster or the measurement uncertainties ($\delta\mathrm{RRM}$). The extrinsic RM consists of components from the IGM, the radio source itself, and an ionospheric component, which are difficult to separate. However, we can estimate the scatter caused by $\mathrm{RM}_\mathrm{extr}$ through sources located far away from clusters. At large radii ($r>3R_\mathrm{500}$), the scatter is expected to be unrelated to the ICM, and caused mainly by the extrinsic scatter between sources and noise scatter due to measurement uncertainties. We find that the standard deviation of the observed RMs corrected for the Galactic contribution is equal to 13 $\pm$ 2 rad m$^{-2}$ (see Fig. \ref{fig:RMvsdistfar}), with the measurement uncertainties accounting for 9 rad m$^{-2}$. This gives an estimate of $\sigma_\mathrm{RM,extr} = 9$ rad m$^{-2}$, which is of the same order as the observed dispersion of extragalactic RMs \citep[e.g.][]{Schnitzeler2010,Vernstrom2019}. 

Using this estimate of the extrinsic scatter, we can define our estimator for the RM scatter induced by the ICM, in the observer frame. We define the RM scatter corrected for measurement errors and extrinsic scatter analogously to \citet[][]{Dolag2001}, as 
\begin{equation}\label{eq:scattercor}
{\sigma}_\mathrm{RRM,cor'} = \sqrt{ {\sigma}_\mathrm{RRM}^2 - \frac{\sum_{i}^{N} (\delta \mathrm{RRM})_i^2}{N-1} - \sigma_\mathrm{RM,extr}^2},
\end{equation}
where $\sigma_\mathrm{RRM}$ is any estimator of the standard deviation, and the sum is taken over all sources $i=1$ to $i=N$ to account for varying measurement errors, $\delta\mathrm{RRM}$, per source, and $\sigma_\mathrm{RM,extr}=9$ rad m$^{-2}$. 


Finally, the cluster rest-frame RM scatter is higher than the observed RM scatter because we artificially measure smaller RMs (by a factor $(1+z)^{-2}$) due to cosmological redshift. This effect is difficult to account for exactly in a stacking experiment but can be approximated by using the median redshift of the cluster sample, weighted by the number of polarised sources per cluster. Finally, we thus define the corrected RRM scatter (standard deviation) as
\begin{equation}\label{eq:scattercor2}
    \sigma_\mathrm{RRM,cor} = \sigma_\mathrm{RM,cor'}  (1+\mathrm{med}[z_\mathrm{cluster}])^2.
\end{equation}
This estimator will be calculated in radial bins to measure the RM scatter profile.


\subsection{Comparison to models}
In the simple scenario of random magnetic field {directions} in cells of size $\Lambda_c$ kpc, which have some {uniform} magnetic field strength and {uniform} electron density, the observed RM is the result of a random walk process. Because of the central limit theorem, the distribution of RMs is then expected to be a Gaussian distribution with zero mean, and variance given by \citep[e.g.][]{Murgia2004}
\begin{equation}\label{eq:sigmaRM_los}
    \sigma^2_\mathrm{RM}(a/R_\mathrm{500}) = 812^2 \Lambda_c \int_\mathrm{LOS} (n_e B_\mathrm{\vert\vert})^2 dl,
\end{equation}
where dl is the infinitesimal path length increment along the line of sight (LOS) in kpc, $n_e$ is measured in cm$^{-3}$ and $B_\mathrm{\vert\vert}$ is the magnetic field strength parallel to the line of sight in $\mu$G. 


In reality, the magnetic field structure will more closely resemble a random field with fluctuations on many spatial scales, and both the magnetic field strength and electron density will scale with radius. Thus comparing observations to more realistic scenarios requires simulated magnetic fields. We followed the approach explained in Section 4 of \citet{Osinga2022} to generate mock rotation measure and depolarisation images for all clusters in our sample which have X-ray observations available (99/124). {We assumed that the clusters have spherical and smooth electron density profiles, with best-fits to the de-projected electron density as a function of radius taken from \citet{AndradeSantos2017}, which are shown in Fig \ref{fig:mean_ne}}. {In reality, there are likely density fluctuations in the ICM. However, these are difficult to measure and on the order of only a few per cent \citep[e.g.][]{Schuecker2004,Churazov2012,Sanders2012}. Magnetic field fluctuations are expected to be stronger. Therefore, we have modelled the magnetic field as a Gaussian random field.}

An example of a mock RM and depolarisation image is shown in Figure \ref{fig:simulation} for a random cluster in our sample. {To compare the mock images to observations, we sample a mock source for every observed source at the same projected radius. If the source was classified as a background source, the integration is done over the full grid, while for a cluster member, the integration is done assuming the source lies at the midplane of the cluster.}

In these models, the magnetic field is assumed to be a three-dimensional Gaussian random field with a single-power law spectrum characterised by the following parameters: $B_0$, $\eta$, $n$, $\Lambda_\mathrm{min}$ and $\Lambda_\mathrm{max}$. The first two denote the variables that parameterise the magnetic field strength, assumed to follow 
\begin{equation}\label{eq:Bvsne}
    B(r) = B_0 \left( \frac{n_\mathrm{e}(r)}{\left< n_\mathrm{e}(0) \right>} \right)^\eta,
\end{equation}
where ${\left< n_\mathrm{th}(0) \right>}$ is the average central electron density of the clusters in our sample, which is equal to $0.056$ cm$^{-3}$. {A relation between $B$ and $n_e$ is predicted by MHD simulations of cluster magnetic fields \citep[e.g.][]{Dolag2005,Vazza2018,Marinacci2018}, with the physical motivation for this relation being related to the fact that magnetic fields and thermal electrons are subject to the same external forces (e.g. turbulence and mergers) or micro-scale plasma processes \citep[e.g.][]{Kunz2011}. Depending on the exact physical conditions in the galaxy cluster, this can give rise to various relations between $B$ and $n_e$. For example, the compression of a tangled magnetic field under the magnetic flux-freezing condition, yields $B \propto n_e^{2/3}$, while assuming that the magnetic energy remains at a constant ratio with respect to the turbulent energy results in $B \propto ne^{1/2}$ \citep[e.g.][and references therein]{Johnson2020}. Observationally, this relation has also been found, with  values of $\eta$ roughly between 0 and 1 \citep[e.g.][]{Bonafede2010,Govoni2017,Stuardi2021}}. {We normalise $B$ to follow Eq. \ref{eq:Bvsne} after generating the Gaussian random field (see \citet[][]{Osinga2022} for details), which technically breaks the null-divergence property of the cluster magnetic field. However, this minimally impacts the final results as found by \citet{Murgia2004} and has been a standard method of generating cluster magnetic fields in the literature \citep[e.g.][]{Murgia2004,Bonafede2010,Vacca2010,Govoni2017,deRubeis2024}.}


The last three parameters encode the power spectrum of the magnetic field:
\begin{equation}
    |B_k|^2 \propto k^{-n},
\end{equation}
between minimum and maximum fluctuations scales that are denoted by $\Lambda_\mathrm{min}$ and $\Lambda_\mathrm{max}$ in image space, respectively. {In the simplified picture of scale-by-scale equipartition between the energy density of magnetic fields and turbulent motions (kinetic energy), we expect Kolmogorov turbulence to result in a magnetic field energy spectrum with} $n=11/3$\footnote{{For a power spectrum that is expressed as a vectorial (3D) form in k-space. The one-dimensional power spectrum would follow $n=5/3$.}} \citep[e.g.][]{Schekochihin2004}. {The assumption of scale-by-scale equipartition is not necessarily valid in all galaxy clusters, and observationally, the spectrum was found to take values between $n=1$ and $n=4$ \citep{Murgia2004,Govoni2006,Guidetti2008,Bonafede2010,Vacca2010,Vacca2012,Govoni2017,Stuardi2021,Osinga2022}}. 
We have computed all models on $1024^3$ pixel grids, to simulate all clusters homogenously. Here, one pixel represents 3 kpc, and clusters are thus simulated out to about $1.5R_\mathrm{500}$, with a minimum fluctuation scale of $\Lambda_\mathrm{min}=6$ kpc. 
These models will be compared to observations in various ways, as detailed in the next section.

\section{Results}\label{sec:results}

\subsection{Average magnetic field strength} \label{sec:avBfield}
\begin{figure}
    \centering
    \includegraphics[width=1.0\columnwidth]{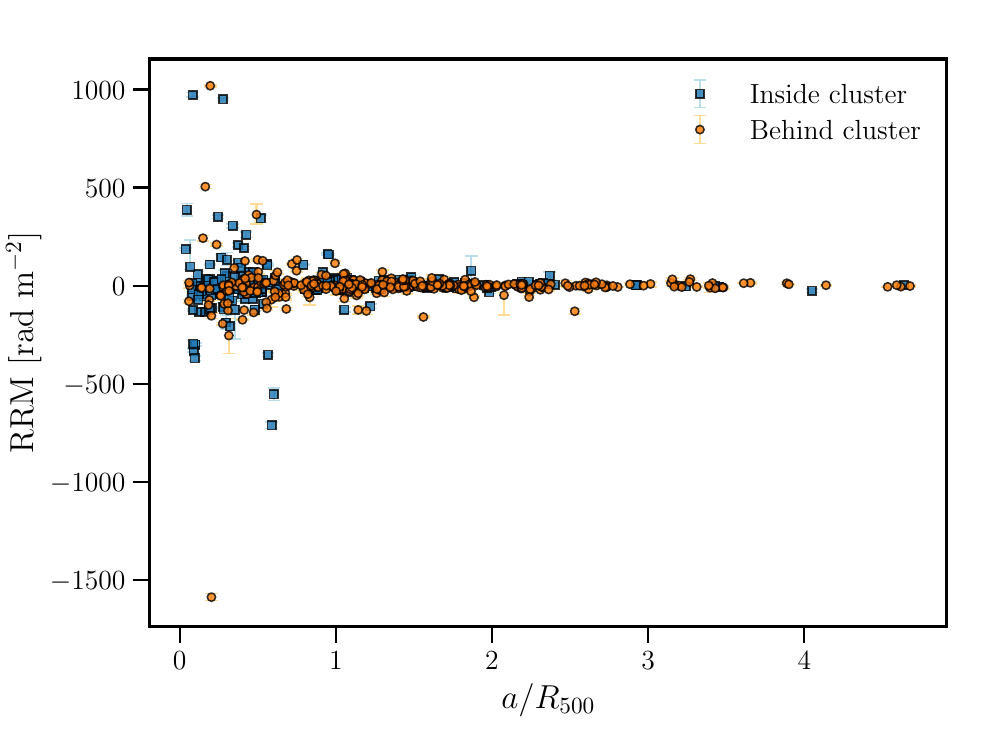}
    \caption{RRM as a function of normalised distance to the nearest cluster centre.  The median uncertainty in RM is 11 rad m$^{-2}$ for both cluster members and background sources. The plot is shown on a partially logarithmic scale in Appendix \ref{appendix:plots} to show low RM sources more clearly.} 
    \label{fig:RMvsdist}
\end{figure}

In Figure \ref{fig:RMvsdist}, we plot the RRM as a function of projected distance to the nearest cluster centre (denoted as $a$ throughout this work). A clear trend is visible, with the scatter in RRM decreasing with a larger distance to the cluster centre. We determined the corrected RRM scatter for all sources within $1R_\mathrm{500}$, using Eq. \ref{eq:scattercor2}, and find that the RRM scatter of cluster members and background sources are similar\footnote{All uncertainties on $\sigma_\mathrm{RRM}$ are calculated by 1\,000 bootstraps.}, being $200 \pm 33$ and $219 \pm 66$ rad m$^{-2}$, respectively, as shown in Table \ref{tab:sigmaRM}. {Assuming that clusters are spherically symmetrical on average, or at least have no bias towards our direction, and that cluster members are on average located at the mid-plane of the cluster, we expect the RRM scatter of background sources to be at least $\sqrt{2}$ times that of cluster members at the same projected radius.} However, cluster members are found preferentially at smaller radii \citep[{e.g.}][{Fig. B1}]{Osinga2022}, where the scatter in RM is larger due to generally larger magnetic field strength and electron densities \citep[e.g.][]{Bonafede2010}. 

\begin{table}[bth]
\caption{Corrected standard deviation of RM in rad m$^{-2}$, as defined in Eq. \ref{eq:scattercor2} for different subsets and projected radii.}
\label{tab:sigmaRM}
\begin{tabular}{@{}lllll@{}}
\toprule
       & $<0.5R_\mathrm{500}$ & $0.5-1.0R_\mathrm{500}$ & $<R_\mathrm{500}$ & $>R_\mathrm{500}$ \\ \midrule
All    & $244\pm47$           & $127\pm30$              & $209\pm37$        & $28\pm5$          \\
Inside\tablefootmark{a} & $196\pm40$           & $219\pm59$              & $200\pm33$        & $27\pm8$          \\
Behind\tablefootmark{b} & $312\pm100$          & $51\pm6$                & $219\pm66$        & $29\pm6$          \\
CC\tablefootmark{c}     & $159\pm66$           & $75\pm24$               & $133\pm49$        & $17\pm3$          \\
NCC\tablefootmark{d}    & $234\pm42$           & $60\pm7$                & $191\pm35$        & $17\pm7$          \\ \bottomrule
\end{tabular}%
\tablefoot{We refer to Fig. \ref{fig:RM_std} for the corrected IQR as a function of radius. \tablefoottext{a}{Sources located inside clusters}\tablefoottext{b}{Sources located behind clusters}\tablefoottext{c}{Only cool-core clusters}\tablefoottext{d}{Only non-cool-core clusters}}
\end{table}

\begin{figure}
    \centering
    \includegraphics[width=1.0\columnwidth]{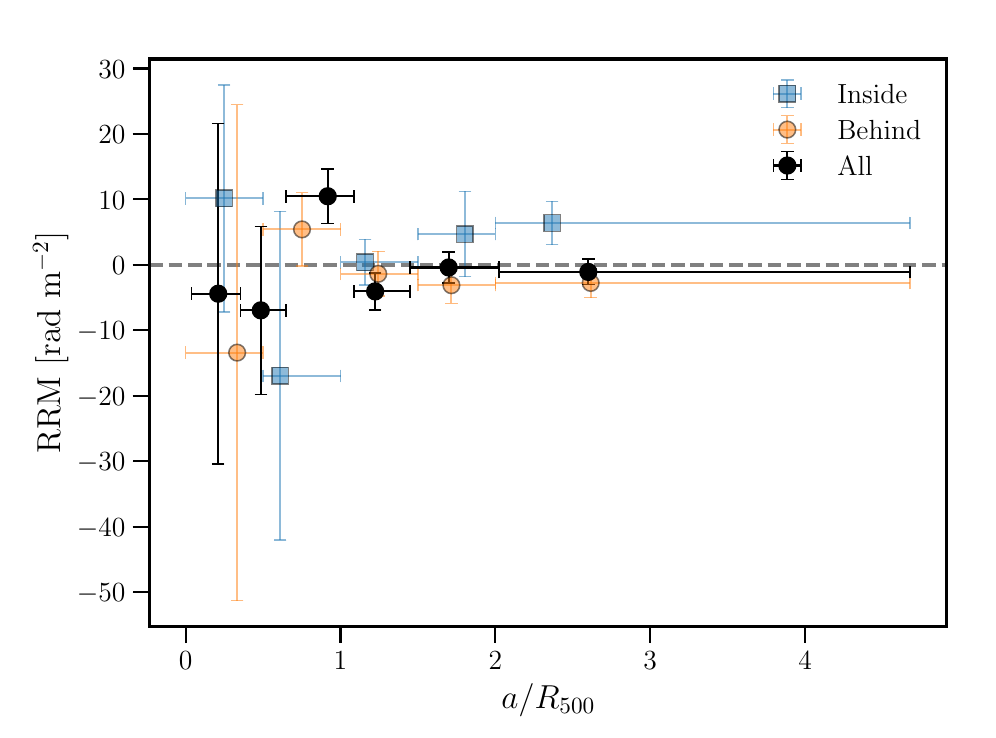}
    \caption{Mean of the RRM in bins of projected normalised distance to the nearest cluster centre. The horizontal error bars indicate the bin edges. Cluster members are shown in blue, background sources in orange, and the combined bins in black.} 
    \label{fig:RMmeanvsdist}
\end{figure}

The mean value of the RMs agrees with zero as a function of radius, as shown in Figure \ref{fig:RMmeanvsdist}, consistent with random magnetic field orientations along the line of sight.  
Assuming the simple random walk scenario denoted by Equation \ref{eq:sigmaRM_los}, we find that the most rudimentary estimate of the line-of-sight magnetic field strength is given by 
\begin{equation}
    \left(\frac{B_\mathrm{\vert\vert}}{\mu\,G}\right)= 2.46 \left(\frac{\sigma_\mathrm{RM}/200}{\,\mathrm{rad \,m}^{-2}}\right) \left(\frac{n_e/10^{-3}}{\,\mathrm{cm }^{-3}}\right)^{-1} \\
    \left(\frac{\Lambda_c/10}{\,\mathrm{kpc}}\right)^{-1/2}
    \left(\frac{L/1000}{\,\mathrm{kpc}}\right)^{-1/2}
\end{equation}
where $L$ indicates the line-of-sight column length, which will be on average twice as large for background sources as cluster members.  If we assume that cells are ordered on scales of $10$ kpc with an electron density of $10^{-3}\, \mathrm{cm}^{-3}$ \citep[e.g.][]{Bohringer2016}, this reduces to
\begin{equation}
    B_\mathrm{\vert\vert} = \frac{\sigma_\mathrm{RM}}{2.57\sqrt{L}}.
\end{equation}
If we assume approximately $L=1000$ kpc for cluster members and twice as large for background sources, we find magnetic field strengths averaged within $R_\mathrm{500}$ equal to 2-3 $\mu$G. Although the simple scenario suffices to give an order of magnitude estimate for the average magnetic field in galaxy clusters, it is clear from Figure \ref{fig:RMvsdist} that the product of the magnetic field strength and electron density is not constant as a function of radius.

\begin{figure}
    \centering
    \includegraphics[width=1.0\columnwidth]{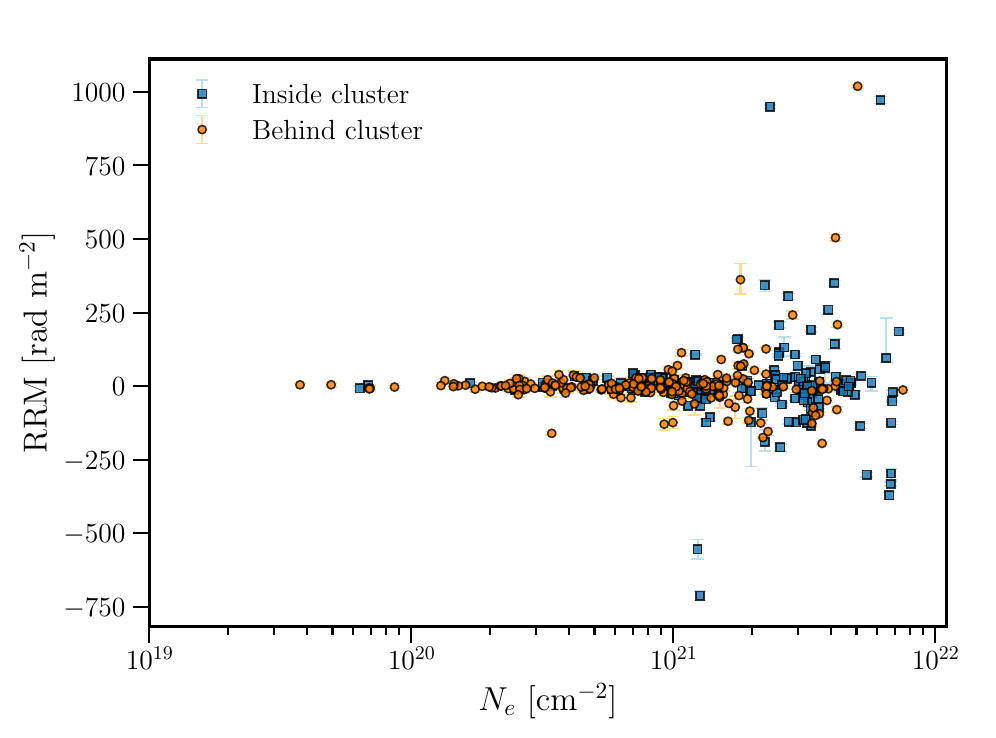}
    \caption{RRM as a function of electron column density, for sources detected at $a<2R_\mathrm{500}$. The plot is shown on a partially logarithmic scale in Appendix \ref{appendix:plots} to show low RM sources more clearly.} 
    \label{fig:RMvscoldens}
\end{figure}

\begin{figure}
    \centering
    \includegraphics[width=1.0\columnwidth]{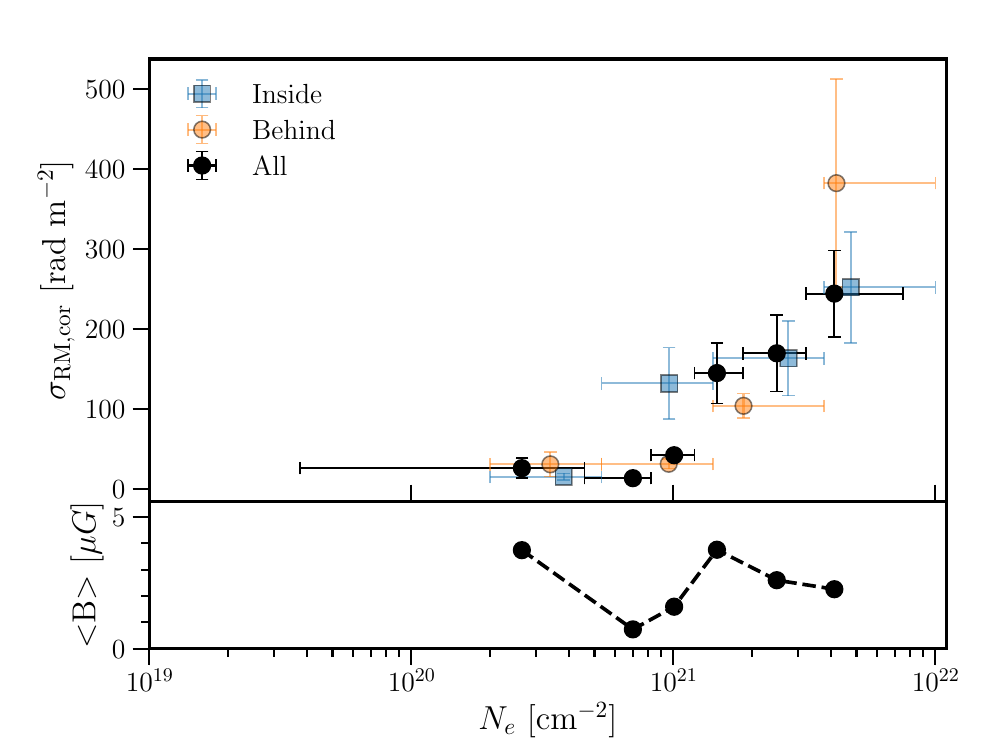}
    \caption{Corrected standard deviation of RMs in bins of electron column density, for sources detected at $a<2R_\mathrm{500}$. The bins for the full sample have equal frequency with 71 sources per bin, while the bins for the sub-samples are defined in logarithmically spaced bins to sample similar densities. The horizontal error bars indicate the bins, and points are plotted at the median $N_e$. The bottom panel shows the magnetic field estimate assuming Equation \ref{eq:BfieldNe}, with a correlation length that is a factor 100 smaller than the line-of-sight distance, resulting in a mean strength of $\sim 2~\mu$G.}
    \label{fig:stdRM_coldens}
\end{figure}

{As we have information on the radial shape of the electron densities}, we plot in Figure \ref{fig:RMvscoldens} the observed rotation measures as a function of \ac{ICM} electron column density. This plot is less populated, as we now only show sources that are detected within a projected radius $a<2R_\mathrm{500}$, where the Chandra-derived column density values are reliable. The scatter in rotation measure significantly increases with increasing column density, and the preferential sampling of cluster members at high column densities (i.e. low radii) is clearly pronounced. Following \citet{Bohringer2016}, we calculated the scatter in rotation measure in bins of column density, from which the average magnetic field strength along the line of sight can be calculated as 
\begin{equation}\label{eq:BfieldNe}
    \left(\frac{B_\mathrm{\vert\vert}}{\mu\,G}\right)= 3.801\times10^{18} \left(\frac{\sigma_\mathrm{RM}}{\,\mathrm{rad \,m}^{-2}}\right) \left(\frac{N_e}{\,\mathrm{cm }^{-2}}\right) 
    \left(\frac{L}{\Lambda_c}\right)^{1/2}.
\end{equation}
The RM scatter as a function of column density is shown in Figure \ref{fig:stdRM_coldens}.
The bottom panel of Figure \ref{fig:stdRM_coldens} shows the resulting magnetic field estimate from Equation \ref{eq:BfieldNe}. We find that the average magnetic field strength is $0.2\mu\mathrm{G} \sqrt{L/\Lambda_c}$, resulting in around $2 \mu$G for typical values of $L=1000$ kpc and $\Lambda_c=10$ kpc \citep{Bohringer2016}. 

\subsection{Radial profile}\label{sec:sigmaRM_radial}
\begin{figure}
    \centering
    \includegraphics[width=1.0\columnwidth]{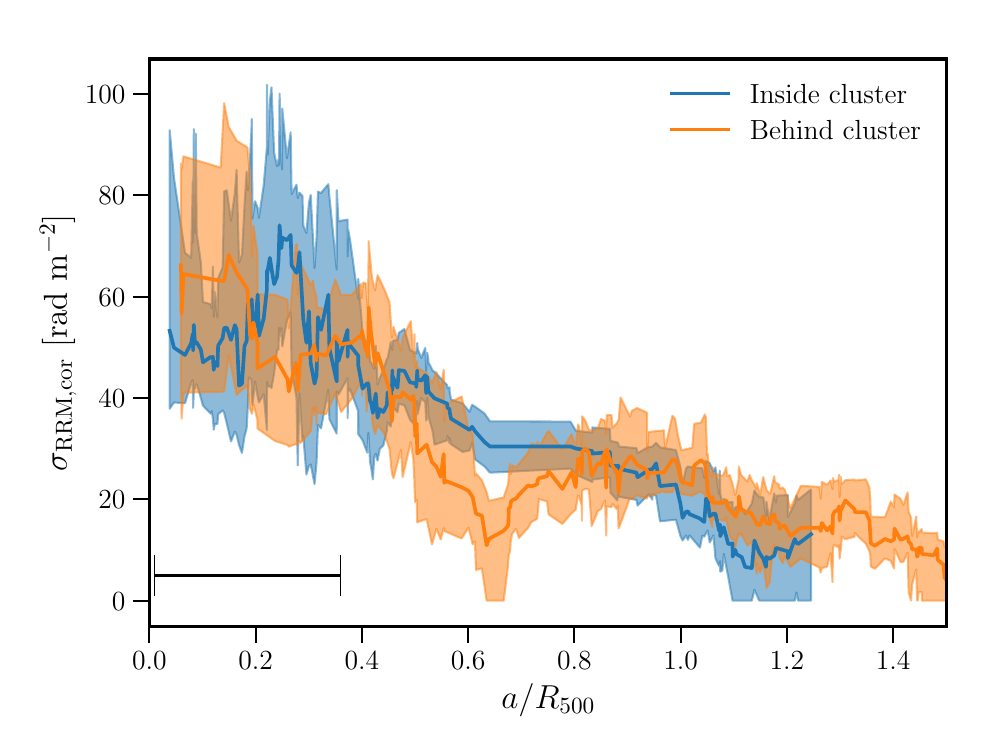}
    \caption{Corrected scatter of observed RRMs as a function of projected distance to the nearest cluster centre. The running scatter is calculated using a fixed sliding window of 50 points, the horizontal error bar indicates the median width of the sliding bin.}
    \label{fig:RM_std}
\end{figure}

Though the average magnetic field strength is a useful property to constrain,  cluster magnetic fields (and electron densities) show a radial decline. We calculated the scatter in rotation measure in a running window of 50 points as a function of radius. This was computed following Eq. \ref{eq:scattercor2}, using the interquartile range (IQR) as a robust measure of  ${\sigma}_\mathrm{RRM}$ (i.e. ${\sigma}_\mathrm{RRM}=\mathrm{IQR}/1.349$). 
A clear trend of the RM scatter being significantly higher closer to the cluster centre is present in Fig. \ref{fig:RMvsdist}, showing the effect of the turbulent magnetised ICM increasing in density, magnetic field strength or both \footnote{We present, in Appendix \ref{app:confounding} an analysis into possible non-ICM effects or confounding variables that could cause such a scaling.}. Background sources and cluster members show similar scatter profiles, indicating that they are largely tracing the same effect. However, background sources do show an increasing scatter above $a\sim 0.7R_\mathrm{500}$. We checked whether this can be attributed to a change in possible confounding variables in Appendix \ref{app:confounding} (i.e. the uncertainty in RRM, galacic latitude of clusters, redshift of clusters, or polarisation fraction of radio sources), and found that this is the region where we start preferentially sampling high-mass clusters, in contrast to the inner region where low-mass clusters are preferentially sampled (see Fig. \ref{fig:cz_vs_rnorm}). As high-mass clusters provide larger column densities, and generally occupy denser parts of the cosmic large-scale structure with higher chances of intervening groups or dense filaments along the line-of-sight, it is likely that this is the cause of the rising RM scatter as a function of radius after $a\sim 0.7R_\mathrm{500}$.

Theoretically, we would expect background sources to display a scatter in RM that is on average greater by a factor $\sqrt{2}$, since cluster members are on average located at the midplane of the cluster and polarised light thus travels through half the column that background sources probe. However, from Figure \ref{fig:RM_std} it is clear that such an effect is not observed.  To quantify this, we performed a z-test to test two null hypotheses in radial bins: i) that the scatter of background sources and cluster members are the same, and ii) that the scatter of background sources is $\sqrt{2}$ times the scatter of cluster members. We found that we could reject neither of the null hypotheses with 95\% confidence. This implies that the uncertainties are too large to identify a statistically significant difference between the RM of background sources and cluster members. To increase the number statistics, we also binned all sources into a radial bin bounded by $[0,R_\mathrm{500}]$, and found marginal evidence ($p=0.02$) to reject the null hypothesis that IQR(RM$_\mathrm{behind})=\sqrt{2}\cdot\,$IQR(RM$_\mathrm{inside})$. However, this is likely caused by the fact that the cluster members are preferentially detected at smaller radii (median radius 0.34$R_\mathrm{500}$) than background sources (median radius 0.54$R_\mathrm{500}$), where the scatter is expected to be larger.

\begin{figure}
    \centering
    \includegraphics[width=1.0\columnwidth]{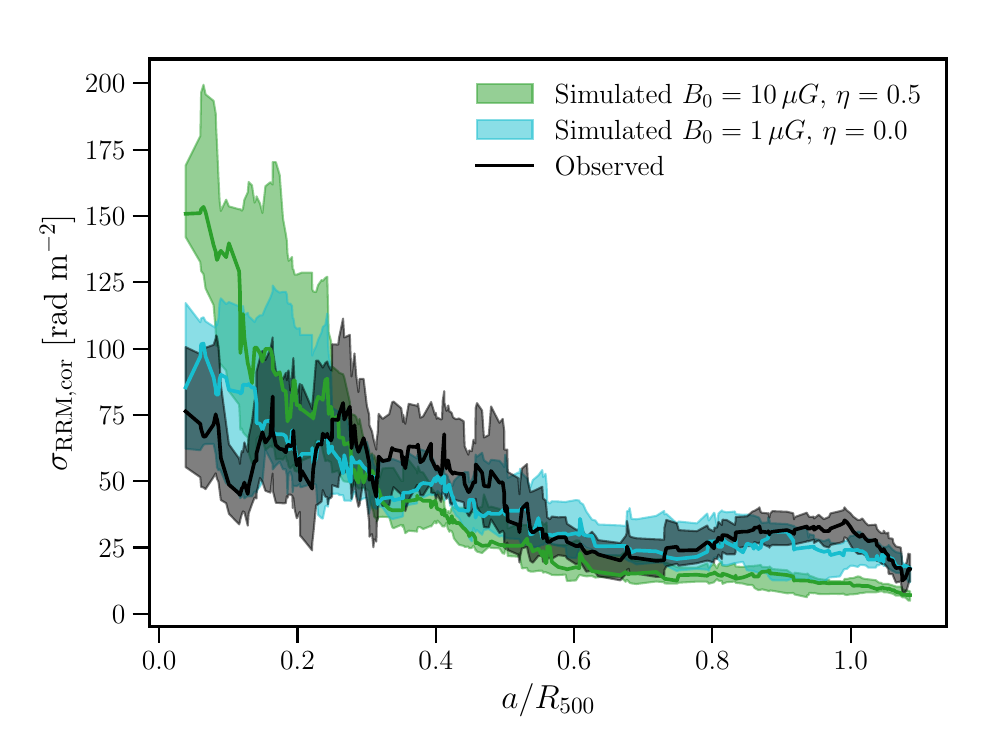}
    \caption{Corrected scatter of RRMs as a function of projected distance to the nearest cluster centre for modelled clusters. The running scatter is calculated using a fixed sliding window of 50 points, with simulated profiles shown for $\eta=0.0$ and $\eta=0.5$.}
    \label{fig:RM_iqr_sim}
\end{figure}

In the simplified model of the turbulent ICM as laid out in Section \ref{sec:methods}, the variance in rotation measure as a function of projected distance should be proportional to the line-of-sight integral given in Equation \ref{eq:sigmaRM_los}. We can thus determine the magnitude of the 3D magnetic field fluctuations and the scaling between magnetic field and thermal electron density from the $\sigma_\mathrm{RRM,cor}$ profile. 
We sampled sources from our mock RM images created with $n=11/3$ (i.e. a Kolmogorov power spectrum), with the same radial sampling as the observed sources. We calculated the simulated scatter profile using the same sliding window, which is shown in Figure \ref{fig:RM_iqr_sim} for a magnetic field to electron density scaling of $\eta=0.5$ and $\eta=0.0$ (i.e. constant $B$). The canonical profile with $\eta=0.5$ is not a good fit to the data, as it underestimates the scatter near the outskirts and overestimates the scatter near the centre.
The observed profile agrees more with a constant magnetic field, although given the smoothing scale of the sliding window, it is difficult to constrain the value of $\eta$ precisely.
The best-fit central magnetic field strength likely lies between $1$ and $10$ $\mu G$, agreeing with the value from the depolarisation of radio sources \citep{Osinga2022}. We note that only using sources behind clusters gives similar results, as the scatter is similar as a function of radius.  However, the relative flatness of the observed profile is caused by the cool-core cluster sample, as investigated in the next section. 


\subsection{Merging vs relaxed clusters}\label{sec:dynstate}
\begin{figure}
    \centering
    \includegraphics[width=1.0\columnwidth]{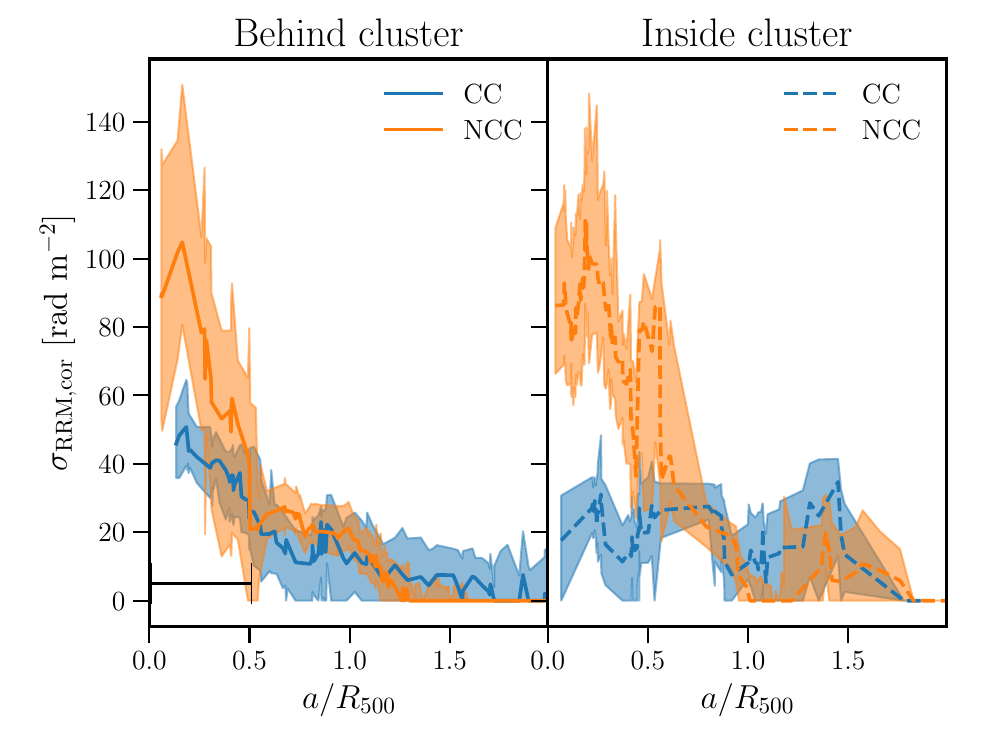}
    \caption{Corrected scatter of RRMs as a function of projected distance to the nearest cluster centre. Cool-core clusters are shown in blue, while non-cool-core clusters are shown in orange. The left panel shows background sources, while the right panel shows cluster members. The running scatter is calculated using a fixed sliding window of size 0.50, indicated by the horizontal errorbar. }
    \label{fig:RM_std_CCvsNCC}
\end{figure}

\begin{figure}
    \centering
    \includegraphics[width=1.0\columnwidth]{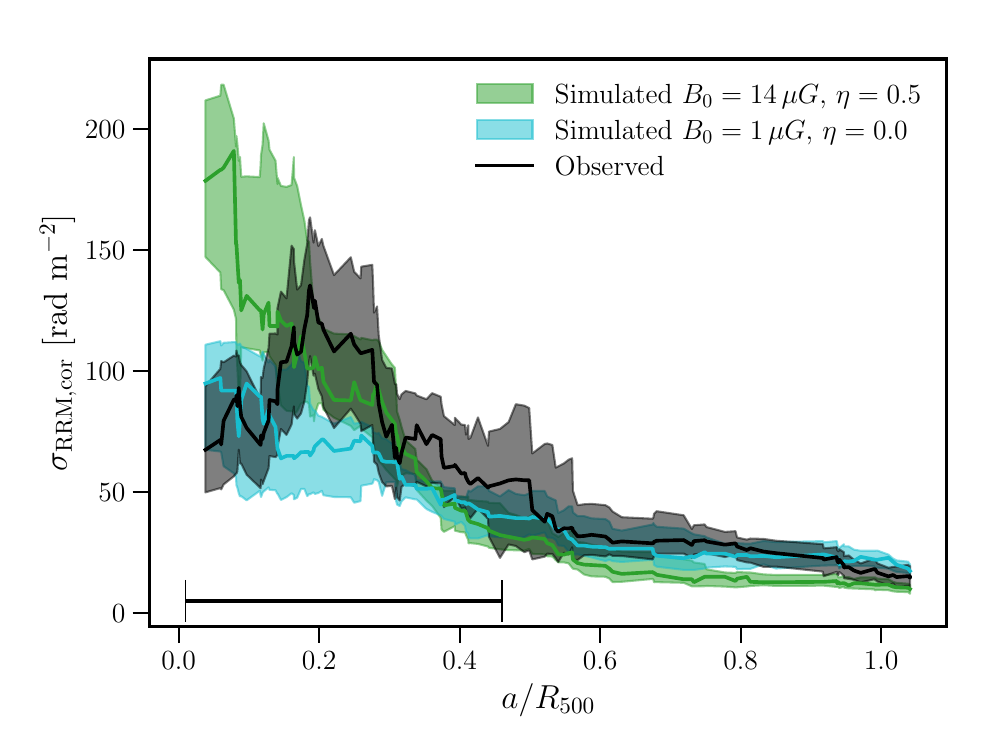}
    \caption{Corrected scatter of RRMs observed in NCC clusters as a function of projected distance to the nearest cluster centre. The running scatter is calculated using a fixed sliding window of 50 points, the median width of the sliding window is indicated by the horizontal error bar. Best-fit simulated profiles are shown for $\eta=0.0$ and $\eta=0.5$.}
    \label{fig:RM_std_NCC_sim}
\end{figure}

Differences between the magnetic field properties of relaxed and merging clusters have been tentatively observed in various studies \citep{Bonafede2010,Stasyszyn2019,Osinga2022}, but not yet clearly quantified. Following \citet{Osinga2022}, we split our sample into merging and relaxed clusters based on the presence of a \ac{CC} or absence of one (NCC). 
The corrected scatter in RM as a function of distance for the NCC and CC clusters is shown in Figure \ref{fig:RM_std_CCvsNCC}. There is a clear difference between the two samples, with the CC sample showing a relatively flat scatter profile, while the NCC sample shows significantly increasing scatter with lower projected radius. Similar behaviour was observed in the depolarisation of radio sources in the same sample, with CC clusters showing a flatter depolarisation profile \citep{Osinga2022}. Additionally, splitting the sample in background sources and cluster members (left panel and right panel) shows that both populations trace similar effects in NCC clusters whereas cluster sources show a significantly flatter profile in CC clusters. It is unlikely that this is an ICM effect, given that it would imply an increasing magnetic field as a function of radius. 
Given the anomalies in the CC sample, 
In Fig. \ref{fig:RM_std_NCC_sim}, we compare the NCC clusters scatter profile to simulated scatter profiles. NCC clusters do show general agreement with a model with $\eta=0.5$ or $\eta=0.0$, although the drop in the central region $a<0.2R_\mathrm{500}$ cannot be explained by our model. This drop is observed in both the background population and sources inside clusters. It is possible that high RRM sources near the projected centre of clusters are depolarised, causing an observational bias against detecting high RRM sources near the centre. We exclude this $a<0.2R_\mathrm{500}$ region in the fit in Fig. \ref{fig:RM_std_NCC_simr02}, where we find that $\eta=0.0$ and $\eta=0.5$ have comparable $\chi^2$ values. 


\subsection{Constraining the magnetic field power spectrum}\label{sec:RMratio}
Both the \ac{RM} and depolarisation of sources scale similarly (although not identically) with the magnetic field strength, but differently with $n$ and $\Lambda_\mathrm{max}$, and therefore the ratio can be used to constrain $n$ and $\Lambda_\mathrm{max}$, relatively independent of the other parameters \citep[e.g.][]{Bonafede2010}. We defined the ratio of RM to depolarisation as 
\begin{equation}\label{eq:RMratio}
    \frac{|\mathrm{RM}|}{1-\mathrm{DP}}.
\end{equation}
This ratio is also relatively independent of radial distance, as the depolarisation will increase (i.e. take lower values of DP) as the RM increases towards the cluster centre. We verified this with both Spearman and Pearson tests, which showed no significant correlation. 

\begin{figure}
    \centering
    \includegraphics[width=1.0\columnwidth]{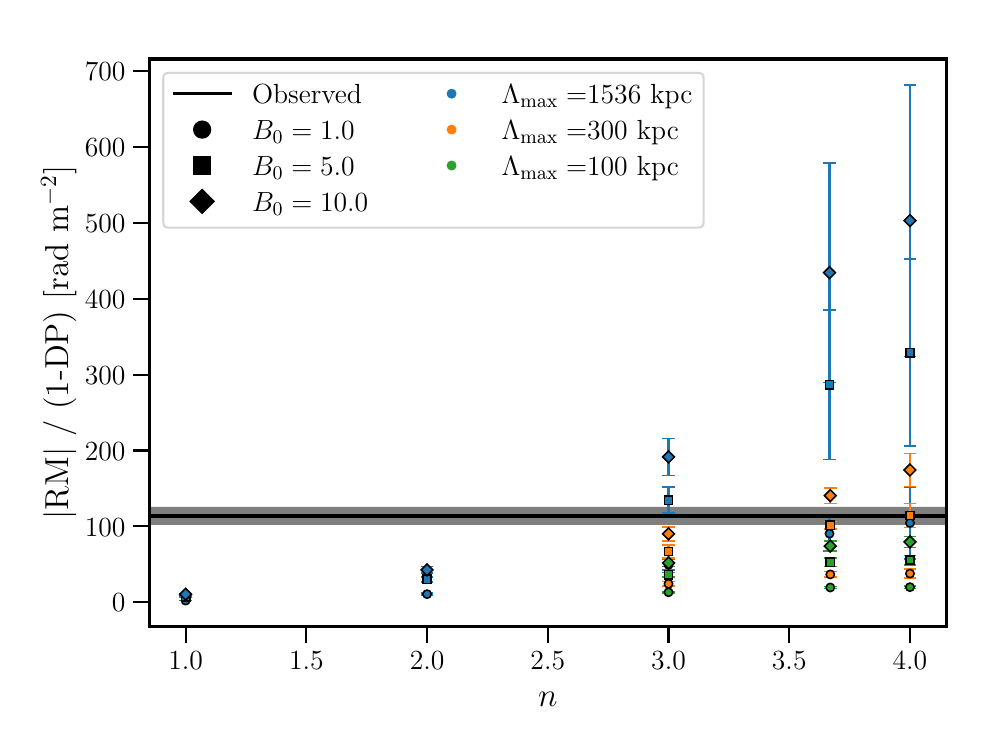}
    \caption{RM ratio, as defined in Equation \ref{eq:RMratio}, as a function of magnetic field power spectrum index $n$. The median observed value is indicated by the black line, while the simulated values are indicated by the coloured points, with different markers for different magnetic field strengths. The uncertainty on the simulated values reflect the 16th and 84th percentile of 10 random initialisations.}
    \label{fig:RMratio}
\end{figure}

The median observed \textbf{ratio of} ${|\mathrm{RM}|}/(1-\mathrm{DP})$ \textbf{across} all sources was found to be $113\pm12$ rad m$^{-2}$, with the uncertainty calculated by $1\,000$ bootstraps, with background sources and cluster members having similar values of $106\pm17$ rad m$^{-2}$ and $119\pm14$ rad m$^{-2}$, respectively. Figure \ref{fig:RMratio} shows the median ratio compared with simulated values for sources sampled at similar positions in the simulated RM and depolarisation images. It is clear from the figure that models with most of the magnetic field energy on small scales (i.e. $n<3$) cannot reproduce the observed RM ratio, mainly because those models result in too low values of $\vert\mathrm{RM}\vert$ due to the rapidly fluctuating magnetic field along the line of sight. Instead, models with $n\geq3$ provide a good fit for various values of the maximum correlation scale $\Lambda_\mathrm{max}$. Lowering $\Lambda_\mathrm{max}$ has an analogous effect to lowering $n$, namely decreasing the coherence length of the magnetic field along the line of sight. This thus results in a smaller average $|\mathrm{RM}|$ while the effect on depolarisation is less significant, as this is measured on scales below the observing beam (less than typically 15 kpc, although dependent on cluster redshift). Thus, the observed data best matches $n\geq3$. 

For a Kolmogorov spectrum ($n=3.67$) with typical values of the central magnetic field strength between $1-10\,\mu$G, the data is consistent with $\Lambda_\mathrm{max}>100$ kpc. Lower values of $\Lambda_\mathrm{max}$ would require significantly higher central magnetic field strengths.

\subsection{RM and depolarisation forward model}\label{sec:forwarmodel}
Finally, we find the best model that reproduces the data by following the approach used in previous works such as \citet{Murgia2004,Bonafede2010,Govoni2017,Stuardi2021}, directly comparing the simulated RM and depolarisation images to the observed data. For every source, we sample an equivalent source from the simulated clusters, and compute the expected RM and depolarisation. We minimize the difference between the simulated and observed radial scatter in RMs, and the depolarisation as a function of radius. We define the function to be minimized, $q$, as follows
\begin{multline}\label{eq:q}
q = q_\mathrm{depol} + q_\mathrm{RM} =  \\ \sum_{r}  \left( \frac{ \mathrm{DP}_\mathrm{obs}(r) - \left< \mathrm{DP}_\mathrm{sim}(r) \right>}{\mathrm{err}(\mathrm{DP}_\mathrm{obs})} \right)^2 
+ 
\left( \frac{ \sigma_\mathrm{RRM,obs}(r) - \left< \sigma_\mathrm{RRM,sim}(r) \right>}{\mathrm{err}(\sigma_\mathrm{RRM,obs})} \right)^2,
\end{multline}
where the observables are calculated in running bins equivalent to Fig. \ref{fig:RM_std}. Simulated observables are averaged over 10 different random initialisations as denoted by the angle brackets. Because sources in this sample were found to be intrinsically depolarised with DP=0.92 at large radii \citep{Osinga2022}, we incorporated this into the simulated depolarisation. We fix the power spectrum to the Kolmogorov value of $n=3.67$ to reduce the computational burden, as Section \ref{sec:RMratio} showed that the RM ratio is consistent only with models that have $n>3$. 

\begin{figure*}
  \begin{subfigure}{0.49\textwidth}
    \includegraphics[width=\linewidth]{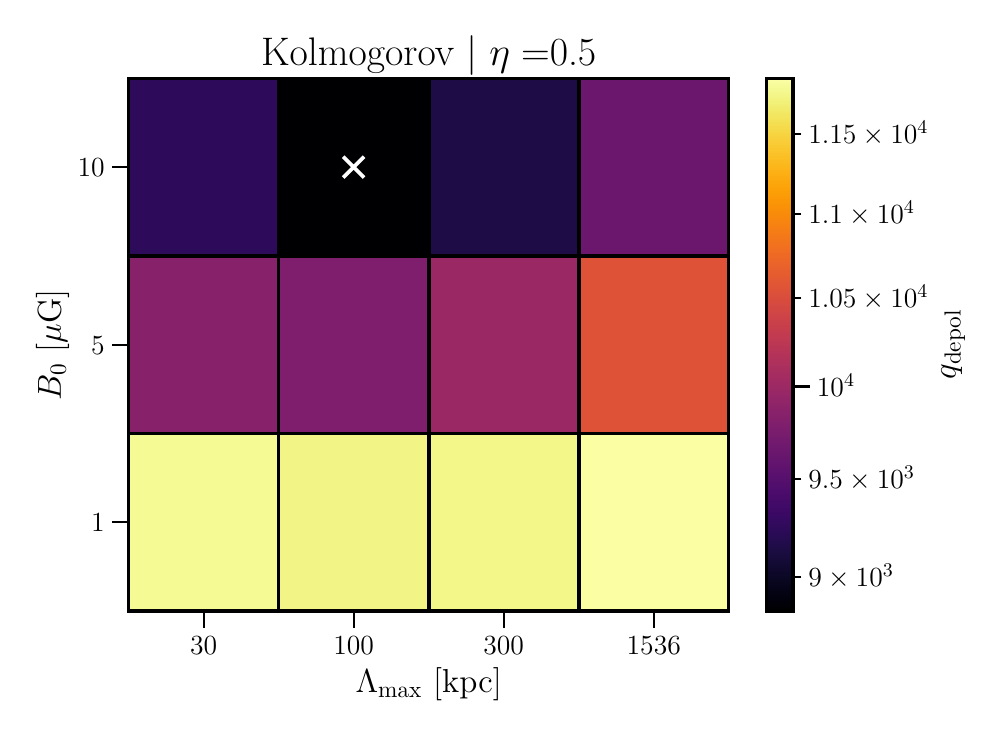}
    \caption{} \label{fig:q1}
  \end{subfigure}%
  \hspace*{\fill}   
  \begin{subfigure}{0.49\textwidth}
    \includegraphics[width=\linewidth]{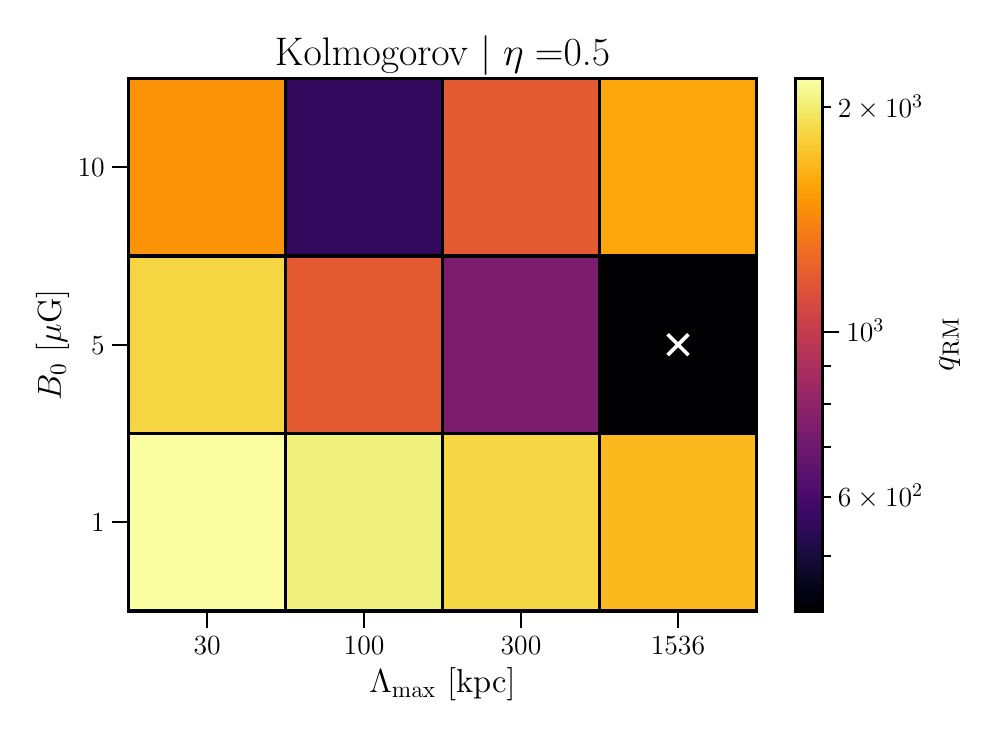}
    \caption{} \label{fig:q2}
  \end{subfigure}%
\caption{Values of $q_\mathrm{depol}$ (a) and $q_\mathrm{RM}$ (b) as defined in Equation \ref{eq:q} for combinations of $B_0$ and $\Lambda_\mathrm{RM}$. Models are simulated with a Kolmogorov power spectrum and $\eta=0.5$. The best-fit model is marked by a cross.} \label{fig:q}
\end{figure*}

\begin{figure*}
  \begin{subfigure}{0.49\textwidth}
    \includegraphics[width=\linewidth]{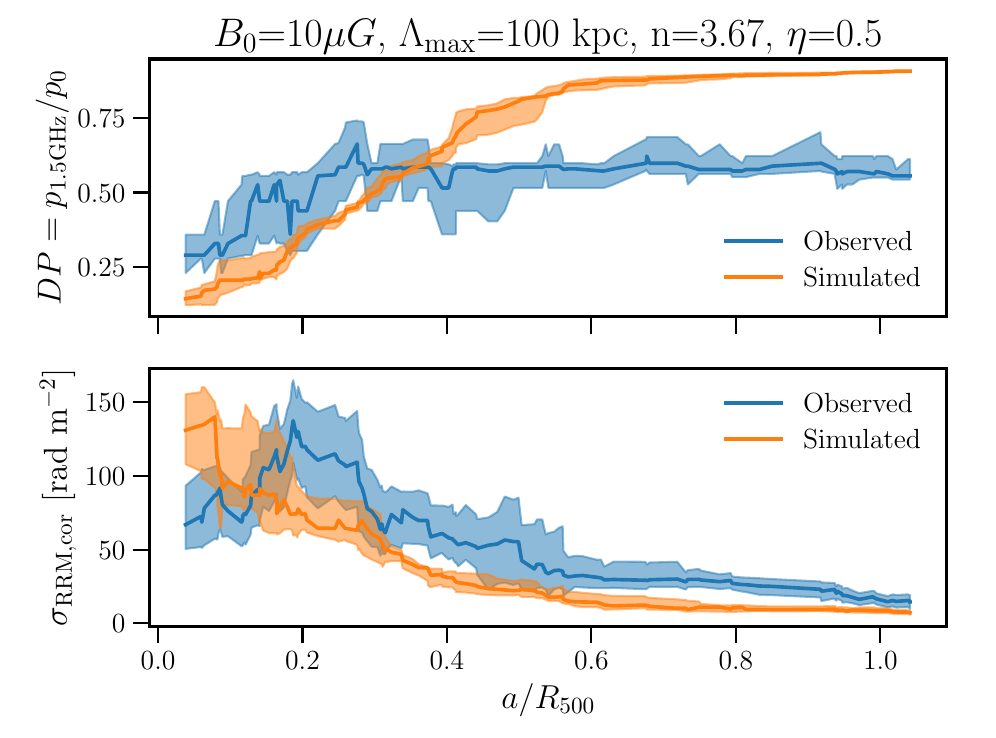}
    \caption{Model that minimizes $q_\mathrm{depol}$.} \label{fig:bestfitKolmogorova}
  \end{subfigure}%
  \hspace*{\fill}   
  \begin{subfigure}{0.49\textwidth}
    \includegraphics[width=\linewidth]{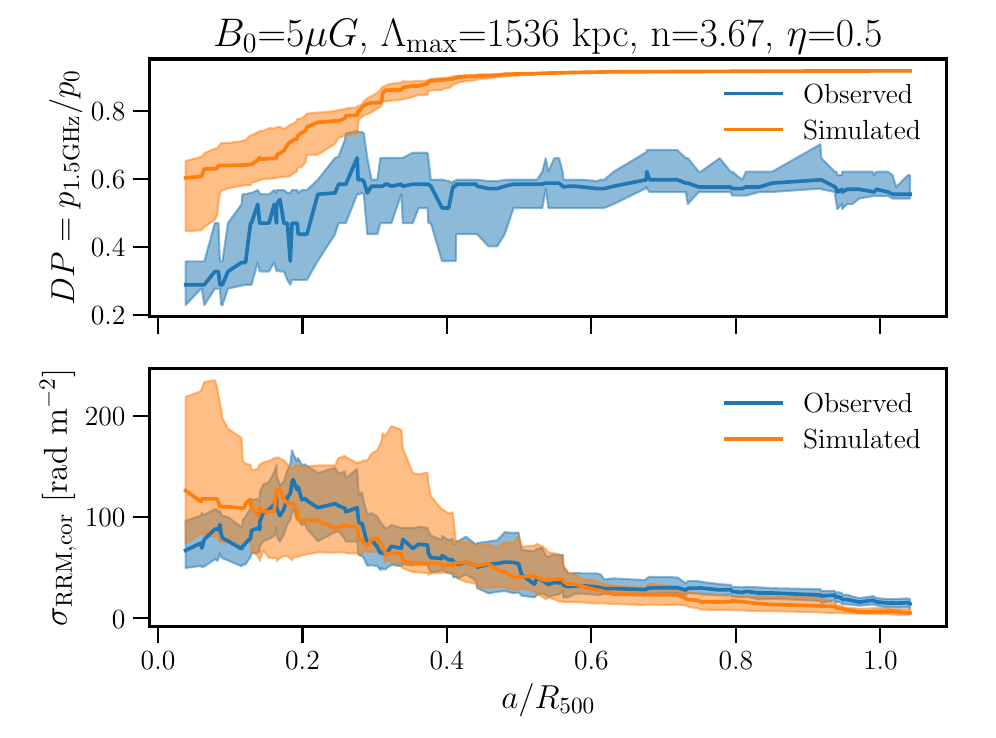}
    \caption{Model that minimizes $q_\mathrm{RM}$.} \label{fig:bestfitKolmogorovb}
  \end{subfigure}%
\caption{Comparison between observed and forward-modelled depolarisation and RM scatter for two models that minimize the q statistic as defined in Equation \ref{eq:q}. } \label{fig:bestfitKolmogorov}
\end{figure*}

Given the atypical behaviour of the CC sample, we forward model only the NCC sample, taking both behind and cluster members since the profiles are statistically similar (see Sec. \ref{sec:dynstate}). This smaller sample size consists of 182 sources detected at $a<1500$ kpc in 42 unique clusters. First, we investigate the best-fit models when fixing the magnetic field to electron density scaling to the typical value of $\eta=0.5$. Figures \ref{fig:q1} and \ref{fig:q2} show the values of $q_\mathrm{depol}$ and $q_\mathrm{RM}$ respectively, as a function of $B_0$ and $\Lambda_\mathrm{max}$. We find the best agreement for a model with $B_0=10\,\mu$G and a maximum correlation scale equal to $\Lambda_\mathrm{max}=100$ kpc for the depolarisation profile and $B_0=5\,\mu$G with $\Lambda_\mathrm{max}=1536$ kpc for the RRM scatter profile. The former is the best-fit model when combining the q values in a normalised form. Figure \ref{fig:bestfitKolmogorov} shows the measured and simulated radial profiles for both best-fit models. Both models are unable to explain the data fully, with both the observed $\sigma_\mathrm{RRM}$ and the depolarisation profile being flatter than the simulated profiles. 

Section \ref{sec:sigmaRM_radial} showed that the radial profile of $\sigma_\mathrm{RRM}$ preferred low values of $\eta \approx 0.0$. We thus also forward modelled with $\eta=0.0$, and show the q-values and best-fit forward models in Appendix \ref{appendix:plots}, Figures \ref{fig:eta00q} and \ref{fig:bestfitKolmogorov_eta00}. Strictly by q-value, models with $\eta=0.0$ provide a better fit to the data than the $\eta=0.5$ models, but again none of the models provide a consistent fit to both the depolarisation and rotation measure scatter that we observed.


\section{Discussion}\label{sec:discussion}
Previous studies of samples of clusters have investigated either the statistical depolarisation \citep{Bonafede2010, Osinga2022} or the rotation measure as a function of projected radius \citep{Clarke2001,Bohringer2016,Stasyszyn2019}. In this work, we have for the first time measured both the rotation measure and depolarisation in a consistent way for polarised radio sources along sight-lines of stacked galaxy clusters. 

The previously largest statistical study of cluster magnetic fields was performed by \citet{Bohringer2016}, who cross-matched 1773 clusters with the RM catalogue compiled by \citet{Kronberg2011} to find 92 RM sight-lines within $R_\mathrm{500}$. We significantly increase the number statistics, detecting 261 radio sources with well-defined RM and depolarisation within $R_\mathrm{500}$ in only 124 clusters. Comparing Table \ref{tab:sigmaRM} with Table 1 in \citet{Bohringer2016}, we find significantly higher $\sigma_\mathrm{RRM}$ in all radial bins, except outside $R_\mathrm{500}$. The reason for this is likely twofold. First, there is a difference in the RM sensitivity as the \citet{Bohringer2016} is likely based on radio data with larger channel widths, preventing detection of high RMs (they find a maximum $\vert \mathrm{RM}\vert<700$ rad m$^{-2}$). Second, the cluster samples are different, as \citet{Bohringer2016} quote a mean cluster mass of $3\times10^{14} M_\odot$, while only 15 out of our 124 clusters have a mass below this value, and the mean mass of our sample is $5.7\times10^{14} M_\odot$. Taking the different sampling into account by calculating the scatter as a function of electron column density in Section \ref{sec:avBfield}, we found an average magnetic field estimate of $3\,\mu$G in our sample of galaxy clusters, perfectly consistent with the findings of \citet{Bohringer2016}. 
Outside the typical $R_\mathrm{500}$ radius of clusters, we observed a corrected standard deviation $\sigma_\mathrm{RRM} = 28 \pm 4$ rad m$^{-2}$. This is lower than the value of $56 \pm 8$ rad m$^{-2}$ found by \citet{Bohringer2016} which did not correct for measurement uncertainties, but significantly higher than the expected intrinsic variation in the radio sources of $5-7$ rad m$^{-2}$ \citep{Schnitzeler2010}. We thus observe an enhanced RM scatter beyond $R_\mathrm{500}$ as well.

Inside $R_\mathrm{500}$, the shape of the scatter profile was used to determine the magnetic field strength and electron density scaling, as shown in Figure \ref{fig:RM_iqr_sim}. The RM scatter inside $R_\mathrm{500}$ showed a flat radial profile compared to expectations from a simple scaling of the magnetic field with the electron density of clusters. Plasma theories generally predict $\eta \geq 0.5$, with magnetic flux freezing giving $\eta=\frac{2}{3}$, adiabatic compression giving $\eta=1.0$ and dynamo models often predicting a constant magnetic energy density to thermal energy density ratio (i.e. $\eta=0.5$). Observationally, the best determined magnetic field profile is that of the Coma cluster, for which $\eta=0.4-0.7$ was found. Other studies using resolved (cluster) radio galaxies also find $\eta\geq0.5$ \citep{Murgia2004,Vacca2012,Govoni2017,Stuardi2021}. 

In contrast, we found that the best-fit $\sigma_\mathrm{RRM}(a)$ model preferred $\eta<0.5$, below the other experimental and theoretical values. However, statistical studies such as this one, that use RMs of unresolved radio sources to determine the RM scatter, suffer from a significant observational bias as sources with high RM values near the centre of clusters are likely to be depolarised and thus missed. The observed electron density profiles, shown in Fig \ref{fig:mean_ne}, illustrate the problem as the electron densities rise strongly towards the core of clusters. Using Eq. \ref{eq:sigmaRM_los}, we expect  $\sigma_\mathrm{RRM}$ to be $\sim\,8$ times larger at $r=0.15R_\mathrm{500}$ than at $r=0.5R_\mathrm{500}$. Since the scatter observed at $0.5R_\mathrm{500}$ is on the order of $50$ rad m$^{-2}$, we expect a scatter on the order of $400$ rad m$^{-2}$ at $r=0.15R_\mathrm{500}$. Sources with such high RM scatter are quickly depolarised at the observed frequencies \citep{Osinga2022}, presenting a missing data problem. This artificially decreases the steepness of the $\sigma_\mathrm{RRM}$ profile, requiring low values of $\eta$ to match the data. In contrast, studies relying on resolved radio sources often probe $\sigma_\mathrm{RRM}$ on smaller scales (i.e. the size of the radio source) where fluctuations are expected to be significantly smaller and thus if a polarised radio source is detected, the scatter can be determined more accurately \citep[e.g.][]{Bonafede2010}. 

Other statistical studies probing many unresolved sightlines such as \citet{Bohringer2016,Stasyszyn2019} have also not observed a strong increase of $\sigma_\mathrm{RRM}$ at low radii, implying the same observational bias. However, at higher frequencies, RM values as high as 10{,}000 rad m$^{-2}$ have been observed in the centre of some clusters\footnote{{Depending on the electron density and magnetic field structure in the cluster, \citet{Taylor1993} determined that a magnetic field strength of 30 $\mu$G was required in case of the Hydra A cluster.}} \citep[e.g.][]{Taylor1993,Baidoo2023}. Higher frequency data that suffers less from depolarisation is thus needed to determine the value of $\eta$ accurately. This will be provided by, for example, the VLA Sky Survey \citep[VLASS;][]{Lacy2020}
Depolarisation does not suffer as strongly from this observational bias at low radii, as upper limits on depolarisation fractions can still be set on sources that are significantly depolarised, given that they are detected at high signal-to-noise in total intensity. However, the depolarisation curves that we observed also do not agree with any of the Gaussian random field models explored in this paper, being generally flatter than models, as also seen in \citep{Osinga2022}.

Thus, we finally fit both the $\sigma_\mathrm{RRM}$ and depolarisation radial profile jointly by means of forward modelling the NCC clusters, since the CC clusters showed anomalous behaviour as explained below. None of the models could explain the RM and depolarisation data consistently, with observations finding a flatter profile than expected from the models. However, our models assume clusters are spherical over-densities without any surrounding large-scale structure, which is unrealistic since clusters will on average lie between one or two filaments \citep[e.g.][]{Pimbblet2004}. The densities and possibly magnetic field strengths in the regions connecting the clusters to the filaments will be higher than our spherical $\beta$-like model, plausibly flattening the observed RM and depolarisation profiles. {Additionally, our models do not account for any fluctuations in the electron density with a similar or different power spectrum than that of the magnetic field. Although there are indications from simulations that the magnetic field and thermal density power spectra could be similar at small scales \citep[e.g.][]{DominguezFernandez2019}, this is difficult to constrain observationally and expected to be on the order of a few per cent \citep[e.g.][]{Churazov2012}. 
Additional density fluctuations would generally increase the modelled RM scatter, and thus require slightly lower magnetic field strengths to fit observations.} {Finally, the assumption that the magnetic fields are Gaussian is in contrast with expectations if the fields are generated by the fluctuation dynamo mechanism, which predicts spatially intermittent fields with non-Gaussian statistics \citep[e.g.][]{Schekochihin2004,Vazza2018,Seta2020PRF,Kopyev2022} }
 Comparisons with cosmological magneto-hydrodynamical (MHD) simulations are an obvious next step, {with realistic turbulent gas profiles, potential non-Gaussian magnetic fields} and clusters that are embedded in the large-scale structure around them \citep[e.g.][]{Marinacci2018}.


Keeping these caveats in mind, we fixed the magnetic field to electron density scaling to the observationally best-determined value of $\eta=0.5$ \citep{Bonafede2011} and found that the best-fit model had $B_0=5\,\mu$G and $\Lambda_\mathrm{max}\sim 100$ kpc. The central magnetic field strength is consistent with previous single object studies as well as statistical studies \citep{Murgia2004,Govoni2006,Guidetti2008,Bonafede2010,Vacca2010,Vacca2012,Govoni2017,Stuardi2021,Bohringer2016}. However, the correlation scale is significantly larger than found in resolved cluster member studies that typically find values around $50$ kpc \citep{Guidetti2008,Bonafede2010,Vacca2012,Govoni2017}, although the resolved sources are often limited in size to less than a few hundred kiloparsec. Studies that use the brightness fluctuations or possibly polarised emission of radio halos to constrain the magnetic field power spectrum found results consistent with outer magnetic field fluctuation scales of $\sim400$ kpc \citep{Govoni2005,Vacca2010}, agreeing with the 100 kpc estimate. Fluctuations on scales of more than a few hundred kpc are also expected theoretically, as the turbulent dynamo process thought to be responsible for magnetic field amplification in galaxy clusters is expected to occur on various scales, from less than a kpc up to a Mpc \citep{Donnert2018}. 



In the forward modelling, we have decided to combine the information from cluster members and background sources to maximise the number statistics, as the $\sigma_\mathrm{RRM}$ profiles are statistically similar, and \citet{Osinga2022} showed that the depolarisation properties of cluster members and background sources are also similar. We checked for a biased contribution to $\sigma_\mathrm{RRM}$ from cluster members in Section \ref{sec:sigmaRM_radial} but found that, using the full sample, we could reject neither the null hypothesis that cluster members and background sources have similar scatter nor the null hypothesis that background sources show a scatter that is $\sqrt{2}$ times larger. This is in accordance with the interpretation from \citet{Osinga2022} that in large samples, cluster members and background sources trace similar ICM effects. However, it is also possible that we do not detect the $\sqrt{2}$ difference because cluster members have slightly enhanced RM scatter due to interaction with the ICM \citep[e.g.][]{Rudnick2003}. Because we could not reject either null hypothesis, a larger number of polarised radio sources is needed to observe whether background sources indeed show an RM scatter that is $\sqrt{2}$ times that of cluster members. 


Finally, we checked for qualitative differences between merging and relaxed clusters. During cluster mergers, turbulence is injected on large scales, which can amplify magnetic fields, and drive large fluctuation scales \citep[e.g.][]{Vacca2018}. Relaxed clusters, on the other hand, are expected to have smaller fluctuation scales, as the energy injected by a previous merger has dissipated through cascades to smaller and smaller scales. 
We found a significant difference between the merging and relaxed cluster samples. Merging clusters show higher values of $\sigma_\mathrm{RRM}$ in the region $0.5<r/R_\mathrm{500}$ than relaxed clusters, and sources inside relaxed clusters probe a flat $\sigma_\mathrm{RRM}$ profile as a function of radius. Higher RM scatter in merging clusters was also observed by \citet{Stasyszyn2019}, although with larger radial bin sizes.
If a scaling between $B$ and $n_e$ is assumed, then CC clusters should show a steeper radial profile near the centre, as CC clusters often have higher central electron densities, as also shown in Fig \ref{fig:mean_ne}. It is unclear what is causing the relative flatness of the $\sigma_\mathrm{RRM}$ profile in the sample of cluster members inside CC clusters, so CC clusters were not used for the full RM and depolarisation forward model. {In the future, exploring a two-component magnetic field model for CC clusters might alleviate the inconsistency, analogous to two-component models that are used to fit CC cluster thermal electron density profiles \citep[e.g.][]{Vikhlinin2006}}.

    



\section{Conclusion}\label{sec:conclusion}
This work has presented the continuation of the study presented by \citet{Osinga2022}, where VLA 1-2 
GHz polarisation observations of 124 massive Planck clusters were presented, and the depolarisation properties of polarised sources were investigated as a function of projected radius to the nearest cluster centre. We have incorporated the additional information from the best-fit RM and constrained cluster magnetic field properties by combining depolarisation and RM in a sample of clusters for the first time. 
We summarise the results of this work as follows:
\begin{enumerate}
    
    \item We have clearly detected the increase of the scatter in rotation measure as a function of decreasing projected radius or increasing electron column density. Averaging all 124 clusters, we find a corrected scatter within $R_\mathrm{500}$ of $\sigma_\mathrm{RM}=209\pm37$ rad m$^{-2}$. The scatter outside of $R_\mathrm{500}$ was found to be $28\pm5$ rad m$^{-2}$, still significantly larger than our measurement uncertainties.
    
    \item Assuming that magnetic fields fluctuate on a single characteristic length scale $\Lambda_c$ with a constant strength, the observed RM scatter agrees with an average magnetic field strength within $R_\mathrm{500}$ equal to $2\,(\Lambda_c/10\mathrm{kpc})^{-0.5}\, \mu$G. 

    \item Differences between the RM scatter of cluster members and background sources are expected from the differences in path lengths alone. However, using the full sample of sources, we could not reject the null hypothesis that cluster members sources show similar $\sigma_\mathrm{RRM}$ profiles as background sources, consistent with the result that background sources and cluster members also show similar depolarisation in the same sample of clusters \citep{Osinga2022}.       
        
    \item The profile of $\sigma_\mathrm{RRM}$ shows a significant decrease as a function of projected radius. However, the profile is flatter than expected from models that have a scaling of $B\propto n_e^{\eta}$ with $\eta=0.5$. We found that CC clusters show a flatter profile than NCC clusters, particularly when traced by cluster members. 

    \item Based on the ratio of RRM and depolarisation, models with magnetic field power spectral indices $n<3$ are strongly rejected by the data. This is consistent with expectations from Kolmogorov turbulence ($n=3.67$)
    
    \item Jointly modelling both the depolarisation and rotation measure of sources in a forward modelling approach, we find that the observations 
    cannot be explained by the simple model of a Gaussian random magnetic field that follows $B\propto n_e^\eta$. {Although $\eta=0$ is slightly preferred over $\eta=0.5$, more advanced modelling of cluster magnetic fields is required.}


\end{enumerate}

In this work, we implicitly assumed that all clusters have the same magnetic field parameters, while in reality, this might be a function of dynamical state, mass, or redshift. The universality of cluster magnetic fields has not been thoroughly tested \citep[e.g.][]{Govoni2017} and might significantly affect our results. Future observations with more sensitive telescopes such as MeerKAT and the Square Kilometre Array could test this assumption by detecting enough polarised sightlines through single clusters such that stacking is not required. {Such observations should take into account that Equation \ref{eq:PvsLamda2} might not hold for all sources, as diffuse radio emission from clusters \citep[e.g.][]{Weeren2019} might induce complexity into the Faraday spectra.}

\section*{Data availability}
Table \ref{tab:polarisedtable} is available in electronic form at the CDS via anonymous ftp to \url{cdsarc.u-strasbg.fr} (\url{130.79.128.5}) or via \url{http://cdsweb.u-strasbg.fr/cgi-bin/qcat?J/A+A/}.
   
\begin{acknowledgements}
We thank the referee, Amit Seta, for a thorough report that has improved the quality of this manuscript. 
      EO and RJvW acknowledge support from the VIDI research programme with project number 639.042.729, which is financed by the Netherlands Organisation for Scientific Research (NWO). RJvW acknowledges support from the ERC Starting Grant ClusterWeb 804208. 
EO thanks both Wout Goesaert and Joppe Swart for the careful looks at his code and Bryan Gaensler for providing comments on the manuscript.
    The Dunlap Institute is funded through an endowment established by the David Dunlap family and the University of Toronto.
    Basic research in radio astronomy at the U.S. Naval Research Laboratory is supported by 6.1 Base Funding.
      This paper has made use of observational material taken with an NRAO instrument. The National Radio Astronomy Observatory is a facility of the National Science Foundation operated under cooperative agreement by Associated Universities, Inc. 
The Pan-STARRS1 Surveys (PS1) and the PS1 public science archive have been made possible through contributions by the Institute for Astronomy, the University of Hawaii, the Pan-STARRS Project Office, the Max-Planck Society and its participating institutes, the Max Planck Institute for Astronomy, Heidelberg and the Max Planck Institute for Extraterrestrial Physics, Garching, The Johns Hopkins University, Durham University, the University of Edinburgh, the Queen's University Belfast, the Harvard-Smithsonian Center for Astrophysics, the Las Cumbres Observatory Global Telescope Network Incorporated, the National Central University of Taiwan, the Space Telescope Science Institute, the National Aeronautics and Space Administration under Grant No. NNX08AR22G issued through the Planetary Science Division of the NASA Science Mission Directorate, the National Science Foundation Grant No. AST-1238877, the University of Maryland, Eotvos Lorand University (ELTE), the Los Alamos National Laboratory, and the Gordon and Betty Moore Foundation. 
    The Legacy Surveys consist of three individual and complementary projects: the Dark Energy Camera Legacy Survey (DECaLS; Proposal ID 2014B-0404; PIs: David Schlegel and Arjun Dey), the Beijing-Arizona Sky Survey (BASS; NOAO Prop. ID 2015A-0801; PIs: Zhou Xu and Xiaohui Fan), and the Mayall z-band Legacy Survey (MzLS; Prop. ID 2016A-0453; PI: Arjun Dey). DECaLS, BASS and MzLS together include data obtained, respectively, at the Blanco telescope, Cerro Tololo Inter-American Observatory, NSF’s NOIRLab; the Bok telescope, Steward Observatory, University of Arizona; and the Mayall telescope, Kitt Peak National Observatory, NOIRLab. The Legacy Surveys project is honoured to be permitted to conduct astronomical research on Iolkam Du’ag (Kitt Peak), a mountain with particular significance to the Tohono O’odham Nation. 
SDSS-IV is managed by the Astrophysical Research Consortium for the Participating Institutions of the SDSS Collaboration including the Brazilian Participation Group, the Carnegie Institution for Science, Carnegie Mellon University, Center for Astrophysics | Harvard \& Smithsonian, the Chilean Participation Group, the French Participation Group, Instituto de Astrof\'isica de Canarias, The Johns Hopkins University, Kavli Institute for the Physics and Mathematics of the Universe (IPMU) / University of Tokyo, the Korean Participation Group, Lawrence Berkeley National Laboratory, Leibniz Institut f\"ur Astrophysik Potsdam (AIP),  Max-Planck-Institut f\"ur Astronomie (MPIA Heidelberg), Max-Planck-Institut f\"ur Astrophysik (MPA Garching), Max-Planck-Institut f\"ur Extraterrestrische Physik (MPE), National Astronomical Observatories of China, New Mexico State University, New York University, University of Notre Dame, Observat\'ario Nacional / MCTI, The Ohio State University, Pennsylvania State University, Shanghai Astronomical Observatory, United Kingdom Participation Group, Universidad Nacional Aut\'onoma de M\'exico, University of Arizona, University of Colorado Boulder, University of Oxford, University of Portsmouth, University of Utah, University of Virginia, University of Washington, University of Wisconsin, Vanderbilt University, and Yale University
This research has made use of the NASA/IPAC extragalactic Database (NED), which is operated by the Jet Propulsion Laboratory, California Institute of Technology, under contract with the National Aeronautics and Space Administration. 
This research has made use of NASA's Astrophysics Data System (ADS). This research made use of Astropy, a community-developed core Python package for Astronomy. This research has made use of ChatGPT to restructure text and code.

\end{acknowledgements}

\bibliographystyle{aa}
\bibliography{firstbib.bib}

\newpage
\begin{appendix}

\FloatBarrier

\section{Confounding variables}\label{app:confounding}
\begin{figure}
	\centering
	\includegraphics[width=1.0\columnwidth]{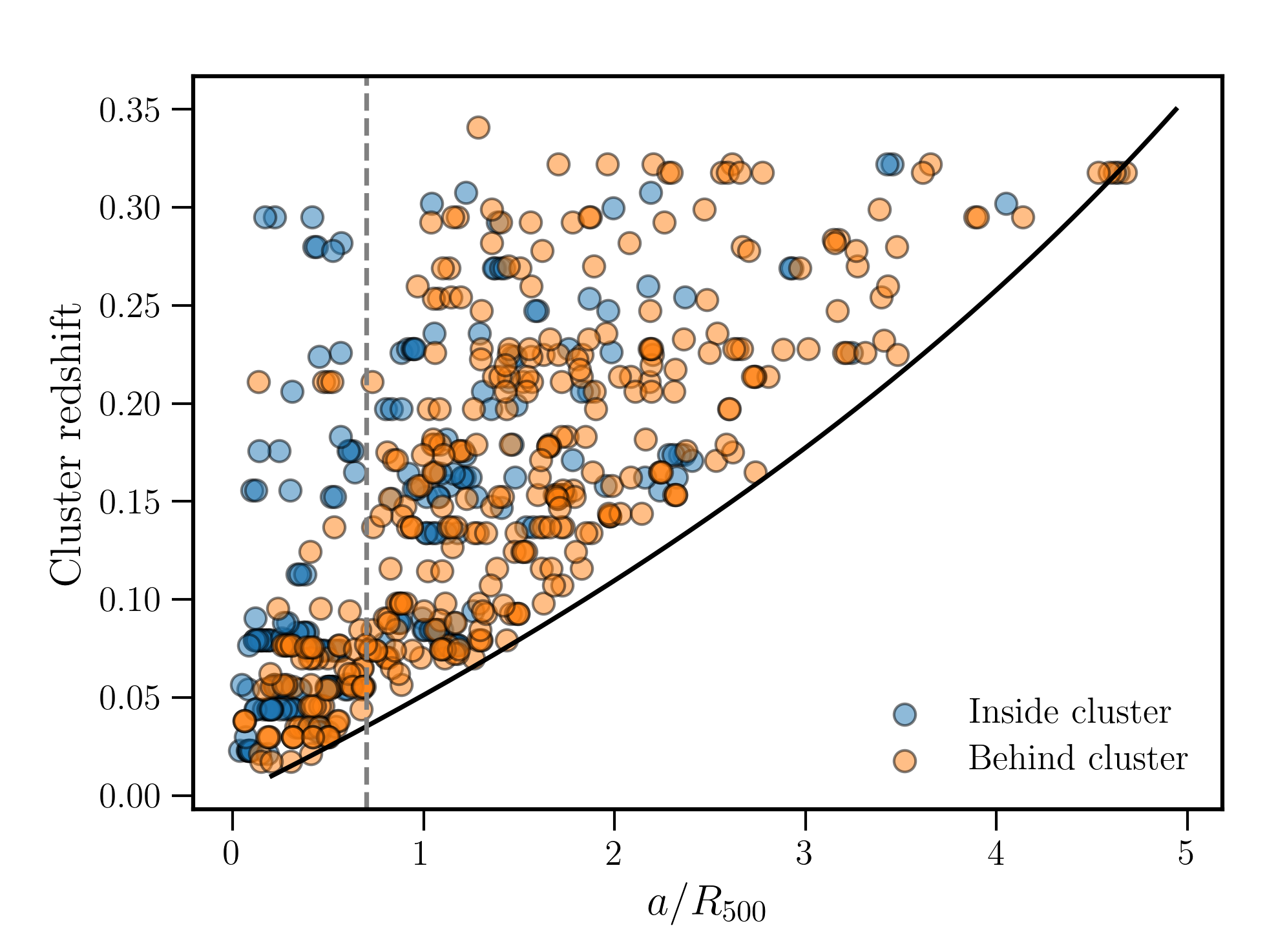}
	\caption{Distribution of all polarised radio sources in the (foreground) cluster vs projected radius plane. The black line shows the FWHM of the primary beam of the VLA L-band observations for a cluster of $R_\mathrm{500}=900$ kpc. The grey line shows the location where background sources start preferentially sampling higher redshift (higher mass) clusters, coinciding with the rising profile in Fig. \ref{fig:RM_std}.} 
	\label{fig:cz_vs_rnorm}
\end{figure}

The radial distribution of polarised sources has an important selection effect. The fact that clusters are observed with pointed observations of a finite primary beam size ($15^\prime$) means that there is a strong correlation between the projected radius of polarised sources and host cluster redshift, as shown in Fig. \ref{fig:cz_vs_rnorm}. This, combined with the fact that PSZ clusters, show a strong correlation between mass and redshift, with lower mass clusters only being found at low redshifts (e.g. Fig. 26 in \citealt{Planck2016}), means that at small radii, the stacked sample is dominated by low-mass clusters, while at large radii the sample is dominated by high-mass clusters. Thus, even if clusters are self-similar in their thermal properties, at larger radii we will be probing preferentially larger column densities and thus expect larger RM scatter and depolarisation. Therefore, a simple analytical evaluation of Eq. \ref{eq:sigmaRM_los} with a fixed path length does not follow the observed trend. However, if the radial profiles are compared to simulated profiles by forward modelling every cluster, integrating along appropriate path lengths, and sampling a simulated radio source at the same radial position, this is taken into account. 

\begin{figure}
	\centering
	\includegraphics[width=1.0\columnwidth]{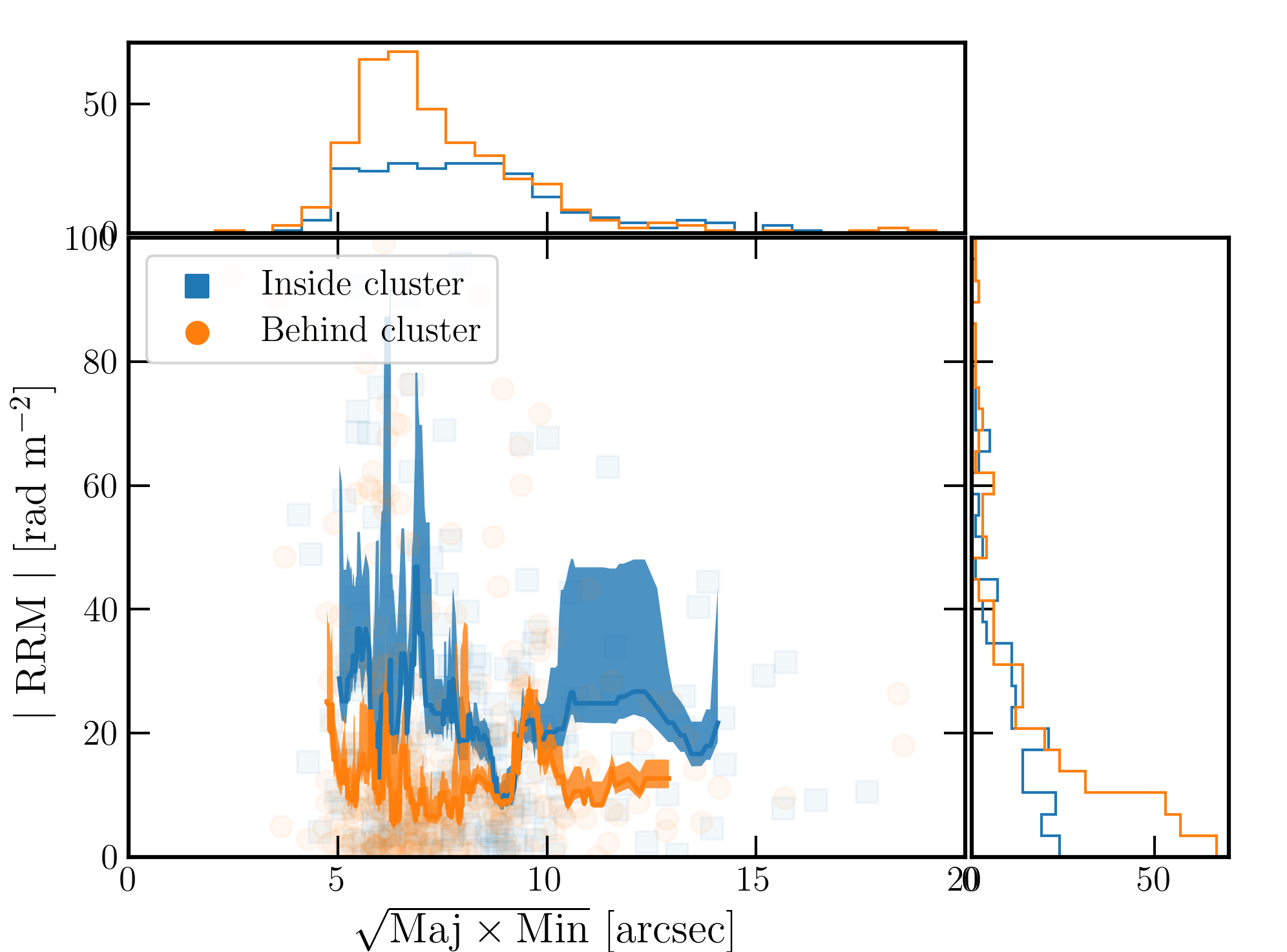}
	\caption{Running median of the absolute value of $\mathrm{RRM}$ as a function of source size. The distribution of sources in this parameter plane is plotted in the background and on the histograms beside the axes.} 
	\label{fig:RRM_vs_size}
\end{figure}

We also investigate whether other variables than the projected radius correlate with increased scatter in RM. We found that the smallest background sources show the largest values of $\vert \mathrm{RRM} \vert$, as shown in Figure \ref{fig:RRM_vs_size}. Particularly, for the 29 sources with $\sqrt{\mathrm{Maj} \times \mathrm{Min}} < 5$ arcseconds, the $\vert \mathrm{RRM} \vert$ rises rapidly. These sources show a median $\vert \mathrm{RRM} \vert = 28$ rad m$^{-2}$, while larger sources show a median of $\vert \mathrm{RRM} \vert = 14.7$ rad m$^{-2}$. This is not fully unexpected, as smaller sources probe a smaller (turbulent) foreground screen, and are thus less affected by smoothing. However, the results are not dominated by these 29 sources, as they are only a small subset of the whole sample, and the radial profile of the scatter is similar when excluding the smallest sources. 

\begin{figure}
	\centering
	\includegraphics[width=1.0\columnwidth]{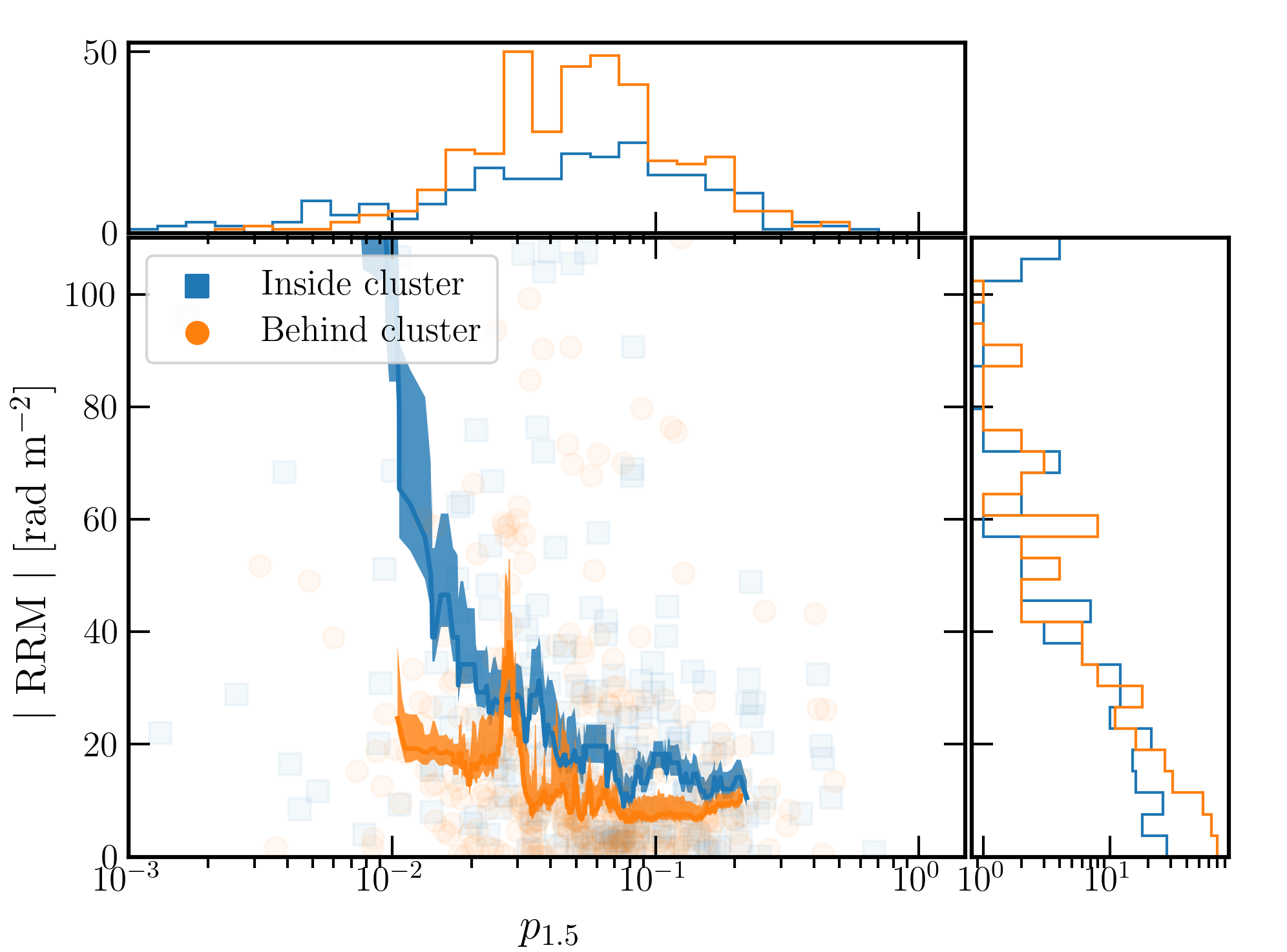}
	\caption{Running median of the absolute value of $\mathrm{RRM}$ as a function of polarisation fraction at 1.5 GHz. The distribution of sources in this parameter plane is plotted in the background and on the histograms beside the axes.} 
	\label{fig:RRM_vs_fracpol}
\end{figure}

\begin{figure}
	\centering
	\includegraphics[width=1.0\columnwidth]{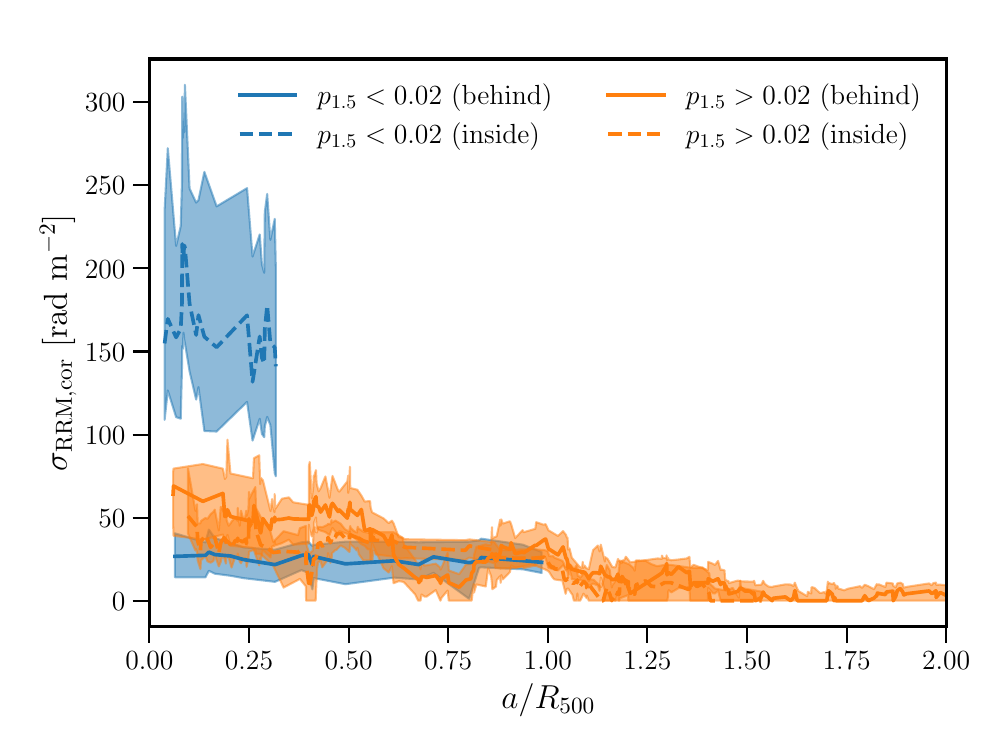}
	\caption{Corrected scatter in RRM as a function of projected radius, separately for sources with low ($p_\mathrm{1.5GHz}<0.02$, 107 sources) and high ($p_\mathrm{1.5GHz}>0.02$, 502 sources) fractional polarisation.} 
	\label{fig:scatter_vs_radius_fracpol}
\end{figure}

There is a strong correlation between the frational polarisation of radio sources and $\vert \mathrm{RRM} \vert$, as shown in Figure \ref{fig:RRM_vs_fracpol}. This effect is particularly strong for cluster sources, which show significantly higher $\vert \mathrm{RRM} \vert $ at low polarisation fractions. Additionally, there are statistically more low polarization sources inside clusters than behind (2\% KS probability of being the same). However, this effect is likely because the strongest depolarisation happens near the centre of the cluster \citep{Osinga2022}, where it is easier to detect cluster members than background sources. It is difficult to separate the effect of an intrinsic relation between low fractional polarisation of sources and higher rotation measures from the assumed effect of the ICM, but we can re-calculate the scatter profile using only sources with a fractional polarisation above 2\%, where the relation between $\vert \mathrm{RRM} \vert $ and $p_\mathrm{1.5GHz}$ flattens. This is shown in Figure \ref{fig:scatter_vs_radius_fracpol}. The high fractional polarisation sources still show decreasing scatter as a function of radius, with similar profiles for cluster members and background sources. It is only the lowest fractional polarisation sources that show a significant difference between cluster members and background sources. Sources with low fractional polarisation are only found inside clusters when they are located near the centre. However, high fractional polarisation sources dominate the analysis, with 502 sources found with $p_\mathrm{1.5GHz}>0.02$ and only 107 sources found with $p_\mathrm{1.5GHz}<0.02$. The profile shown in Fig. 6 is thus dominated by high fractional polarisation sources, which explains that cluster and background sources follow similar profiles. 

Finally, there is a strong correlation between $\vert \mathrm{RRM} \vert$ and Stokes I flux density for cluster members, where brighter sources in total intensity preferentially show higher RRM. \citet{Osinga2022} showed that we preferentially sample brighter sources near the cluster centre (i.e. there is a correlation between total flux density and projected radius from the cluster centre for cluster radio sources), which explains why the most centrally located, and thus brightest sources, correlate positively with $\vert \mathrm{RRM} \vert$. 
In conclusion, the aforementioned correlations are either not dominating the observed trend (i.e. in the case of small sources, or low fractional polarisation) or are likely a by-product of the observed ICM effect (i.e. in the case of the flux density correlation).

\begin{figure}
	\centering
	\includegraphics[width=1.0\columnwidth]{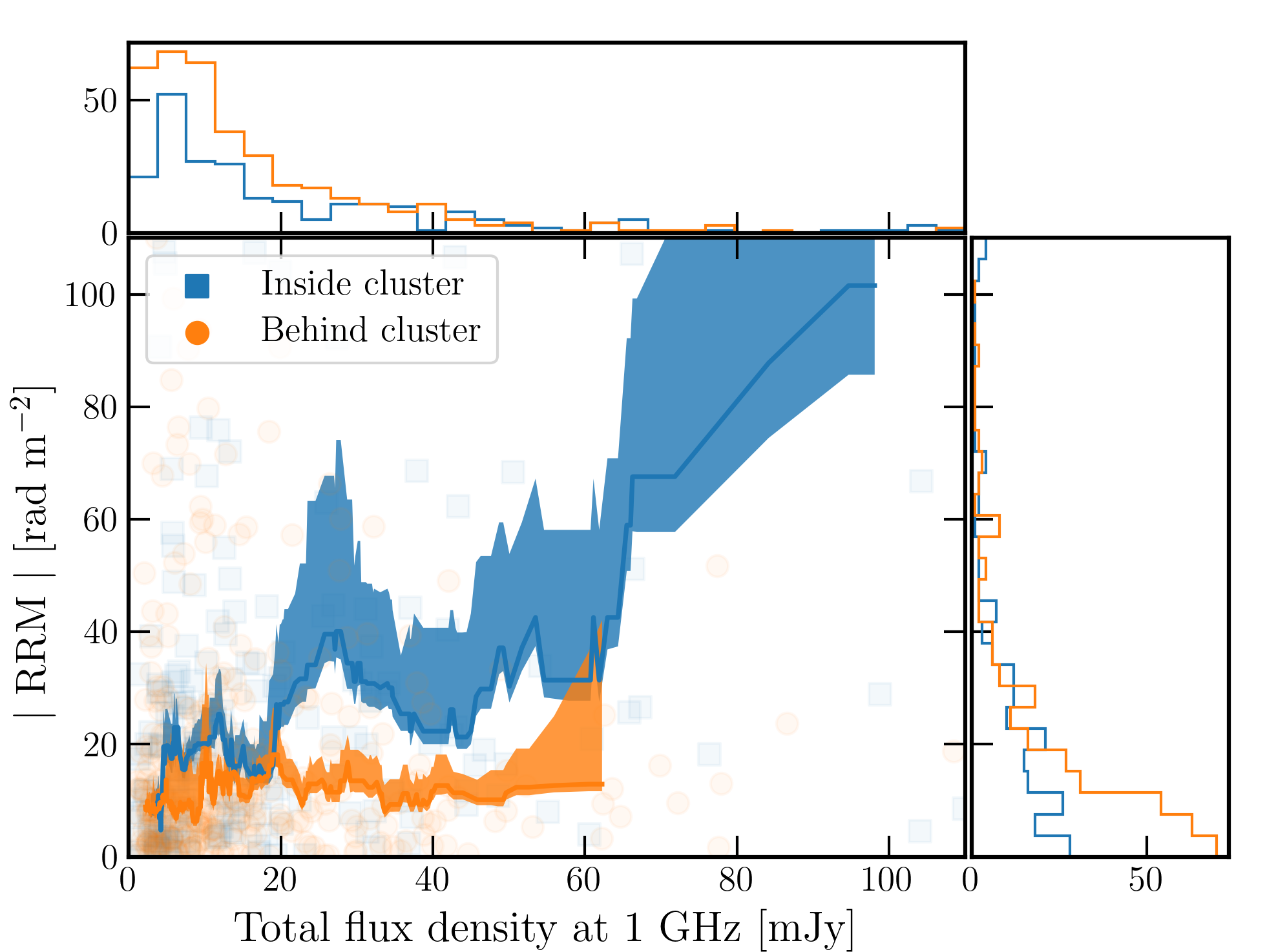}
	\caption{Absolute value of RRM as a function of Stokes I flux density.} 
	\label{fig:abs_RM_vs_I0}
\end{figure}

\FloatBarrier

\section{Alternative visualisations}\label{appendix:plots}
\begin{figure}[htb]
    \centering
    \includegraphics[width=0.85\columnwidth]{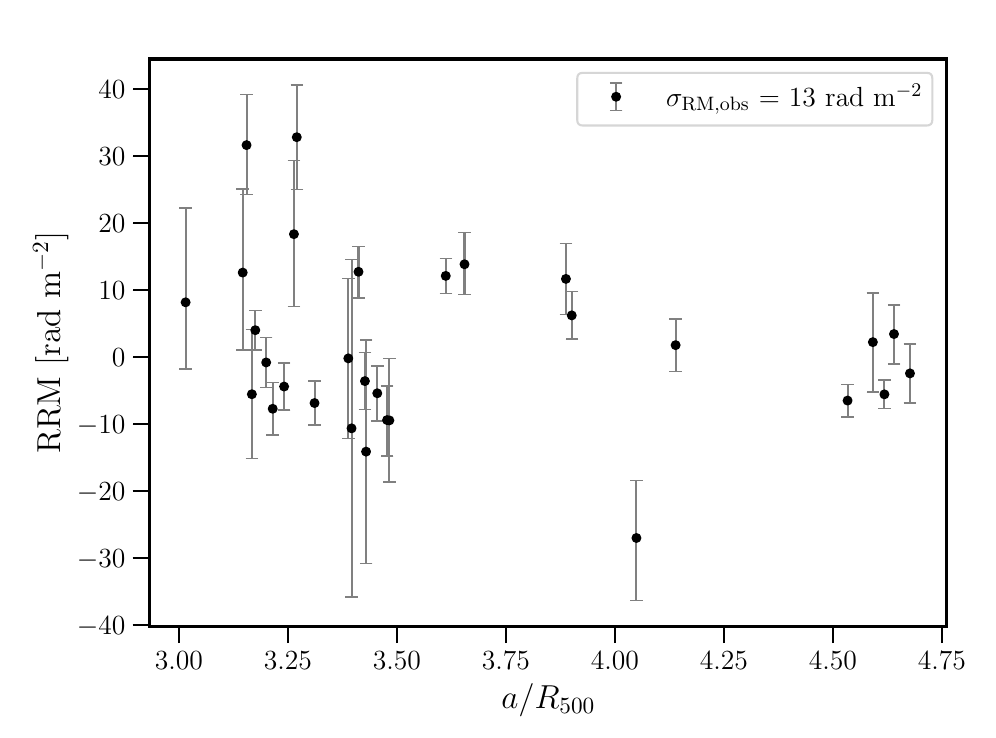}
    \caption{$\mathrm{RM}_\mathrm{obs'}$ as a function of normalised distance to the nearest cluster centre for sources located far away from clusters.} 
    \label{fig:RMvsdistfar}
\end{figure}

In Fig. \ref{fig:RMvsdistfar} we show a zoomed-in version of Fig. \ref{fig:RMvsdist} beyond $a=3R_\mathrm{500}$, used to determine the scatter between sources far outside clusters. {We find no significant correlation between RRM and projected radius for sources far outside the central regions, although statistically significant peaks in RRM have been found at large radii in a larger sample of RRMs around cosmic overdensities \citep{Anderson2024}}.

\begin{figure}[htb]
    \centering
    \includegraphics[width=0.85\columnwidth]{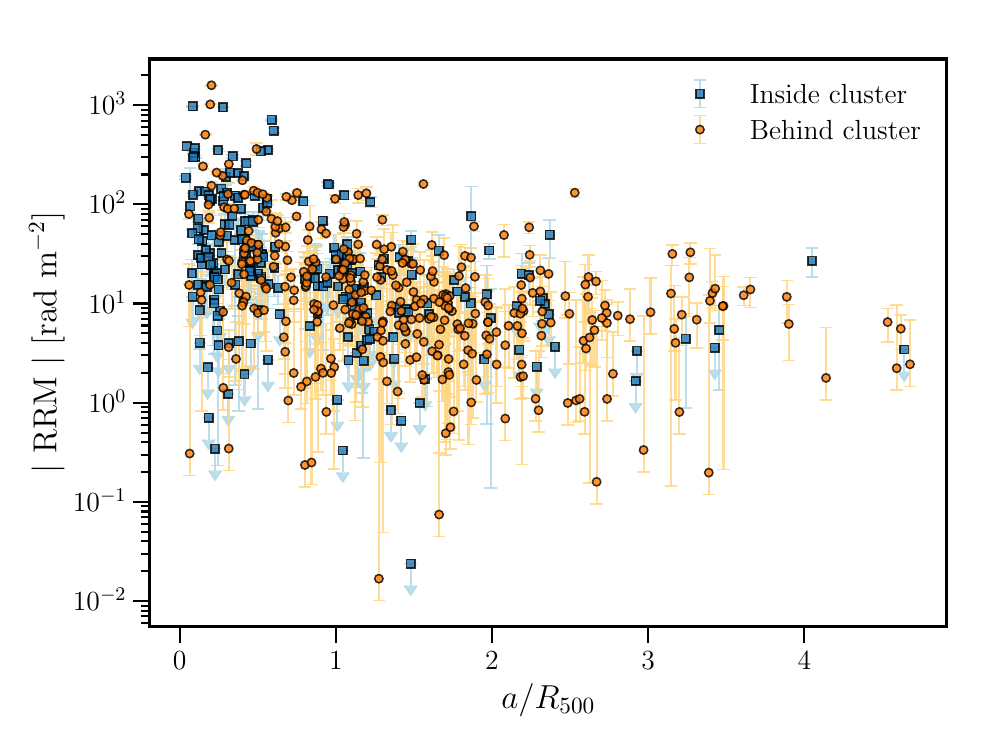}
    \caption{RRM as a function of normalised distance to the nearest cluster centre. The y-axis is logarithmic for visualising low RM sources better than in Fig. \ref{fig:RM_std}. Uncertainties that are consistent with zero RM are plotted as downward-facing arrows.} 
    \label{fig:RMvsdist_log}
\end{figure}

\begin{figure}[htb]
    \centering
    \includegraphics[width=0.85\columnwidth]{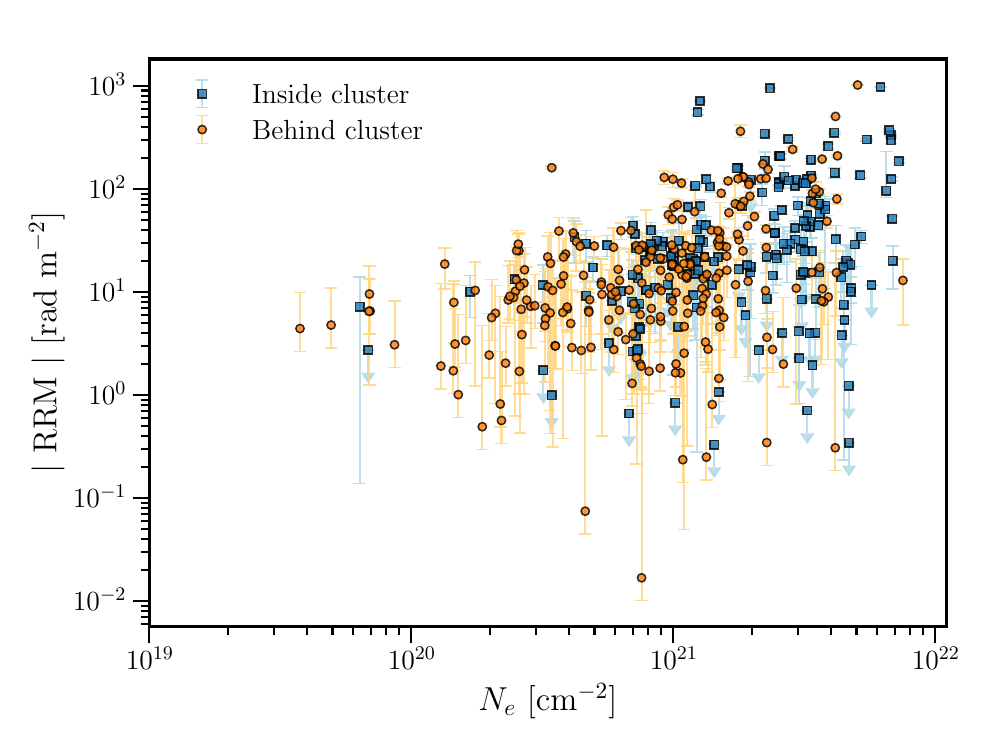}
    \caption{RRM as a function of electron column density. The y-axis is logarithmic for visualising low RM sources better than in Fig. \ref{fig:RMvsdist}. Uncertainties that are consistent with zero RM are plotted as downward-facing arrows.} 
    \label{fig:RMvscoldens_lin}
\end{figure}

\begin{figure}[htb]
    \centering
    \includegraphics[width=0.85\columnwidth]{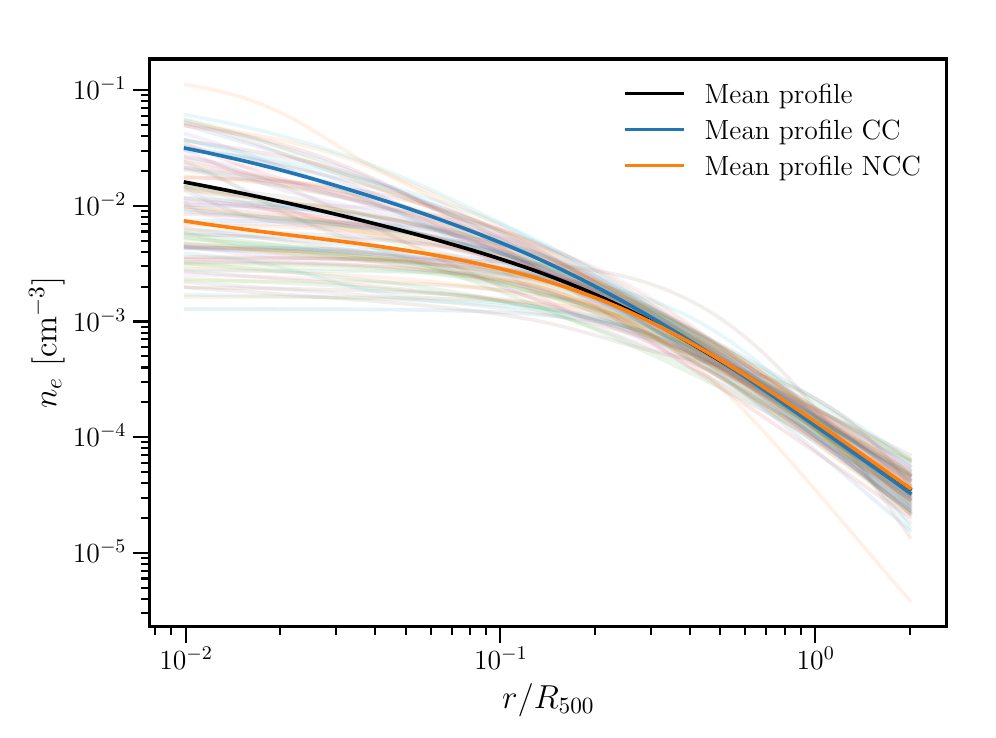}
    \caption{Average electron density profile as a function of normalised 3D radius for the 92 clusters that have Chandra X-ray observations. The mean profile is shown in the black line.} 
    \label{fig:mean_ne}
\end{figure}

To better visualise the low RM sources in our sample, we plot in Figures \ref{fig:RMvsdist_log} and \ref{fig:RMvscoldens_lin} the cluster-induced rotation measure on a logarithmic scale as a function of projected radius and column density, respectively. The electron density profiles of the clusters in our sample, as determined in \citet{AndradeSantos2017}, are shown in Figure \ref{fig:mean_ne}.
Finally, the $q$ values of the full forward model for $\eta=0.0$ are shown in \ref{fig:eta00q}. The best-fit profiles for these fits are shown in Fig \ref{fig:bestfitKolmogorov_eta00}. 

\FloatBarrier

\FloatBarrier

\begin{figure}[htb]
    \centering
    \includegraphics[width=0.85\columnwidth]{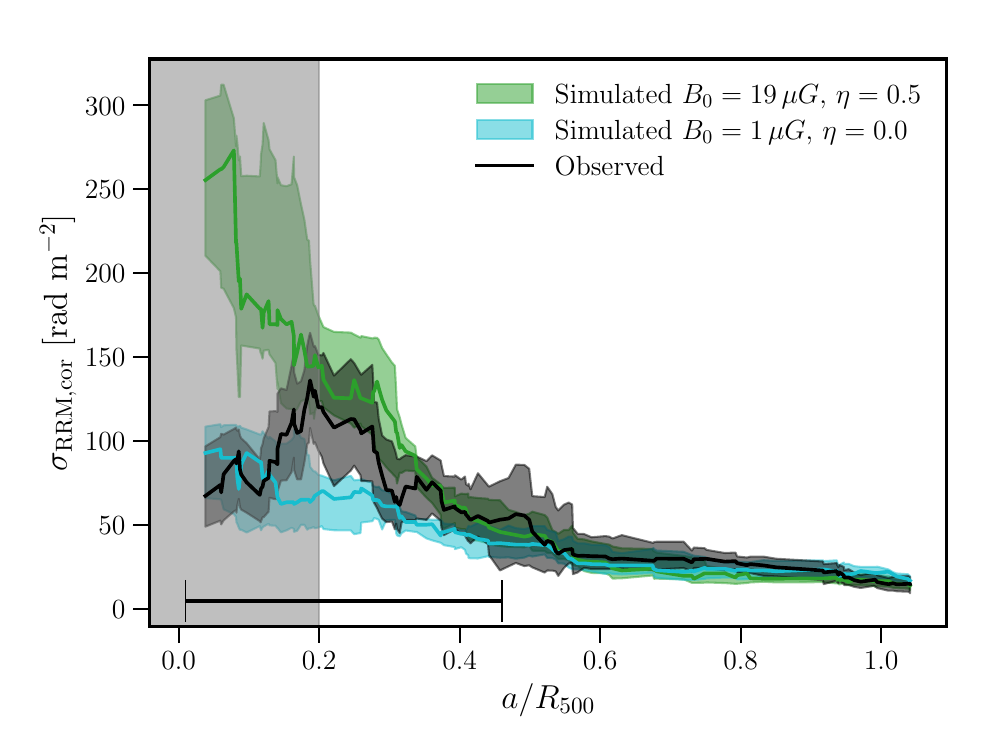}
    \caption{Like Fig. \ref{fig:RM_std_NCC_sim} but fit using only the data at $r>0.2R_\mathrm{500}$.} 
    \label{fig:RM_std_NCC_simr02}
\end{figure}

\begin{figure*}
  \begin{subfigure}{0.49\textwidth}
    \includegraphics[width=\linewidth]{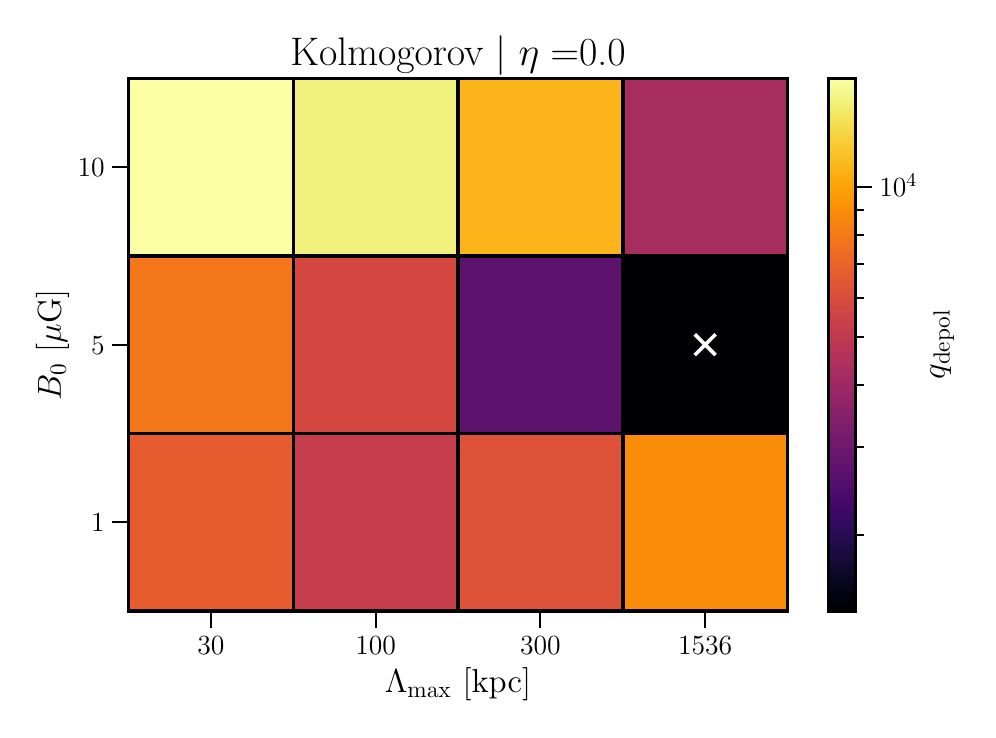}
    \caption{} \label{fig:q1eta01}
  \end{subfigure}%
  \hspace*{\fill}   
  \begin{subfigure}{0.49\textwidth}
        \includegraphics[width=\linewidth]{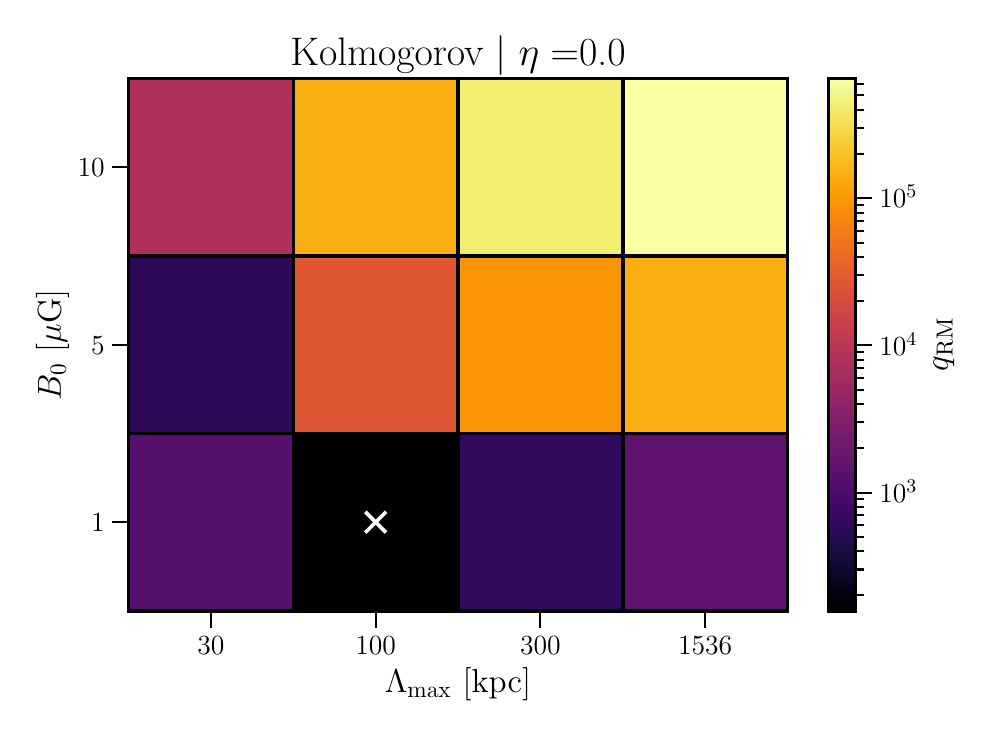}
    \caption{} \label{fig:q2eta01}
  \end{subfigure}%
\caption{Values of $q_\mathrm{depol}$ (a) and $q_\mathrm{RM}$ (b) as defined in Equation \ref{eq:q} for combinations of $B_0$ and $\Lambda_\mathrm{RM}$. Models are simulated with a Kolmogorov power spectrum and $\eta=0.0$. The best-fit model is marked by a cross.} \label{fig:eta00q}
\end{figure*}

\begin{figure*}
  \begin{subfigure}{0.49\textwidth}
    \includegraphics[width=\linewidth]{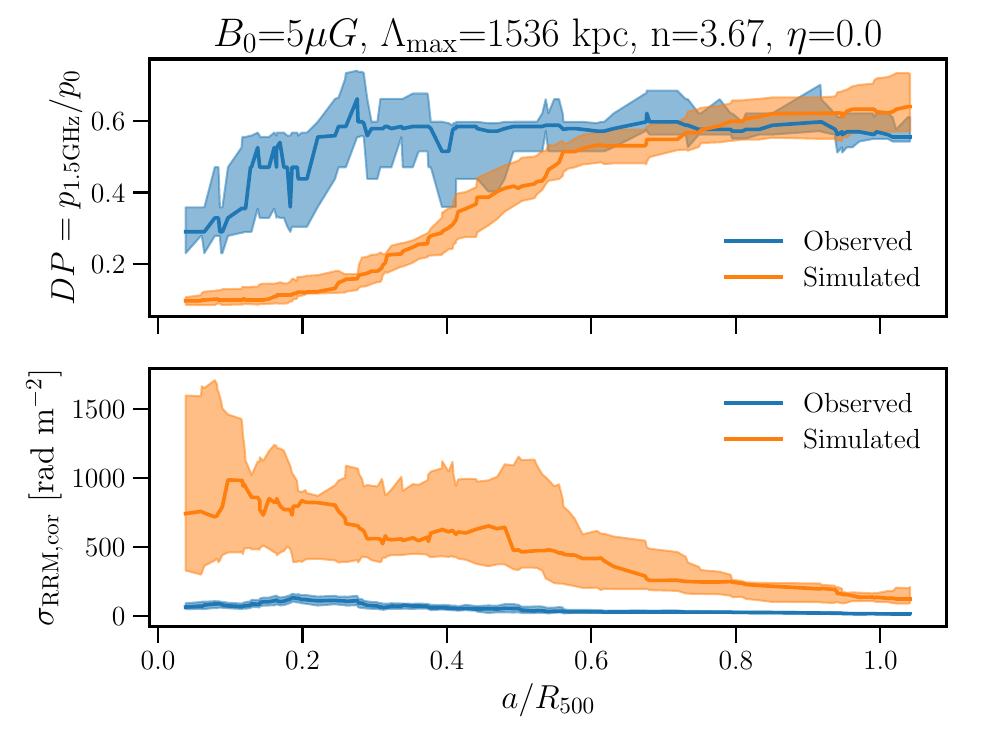}
    \caption{Model that minimizes $q_\mathrm{depol}$.} \label{fig:bestfitKolmogorov_eta00a}
  \end{subfigure}%
  \hspace*{\fill}   
  \begin{subfigure}{0.49\textwidth}
    \includegraphics[width=\linewidth]{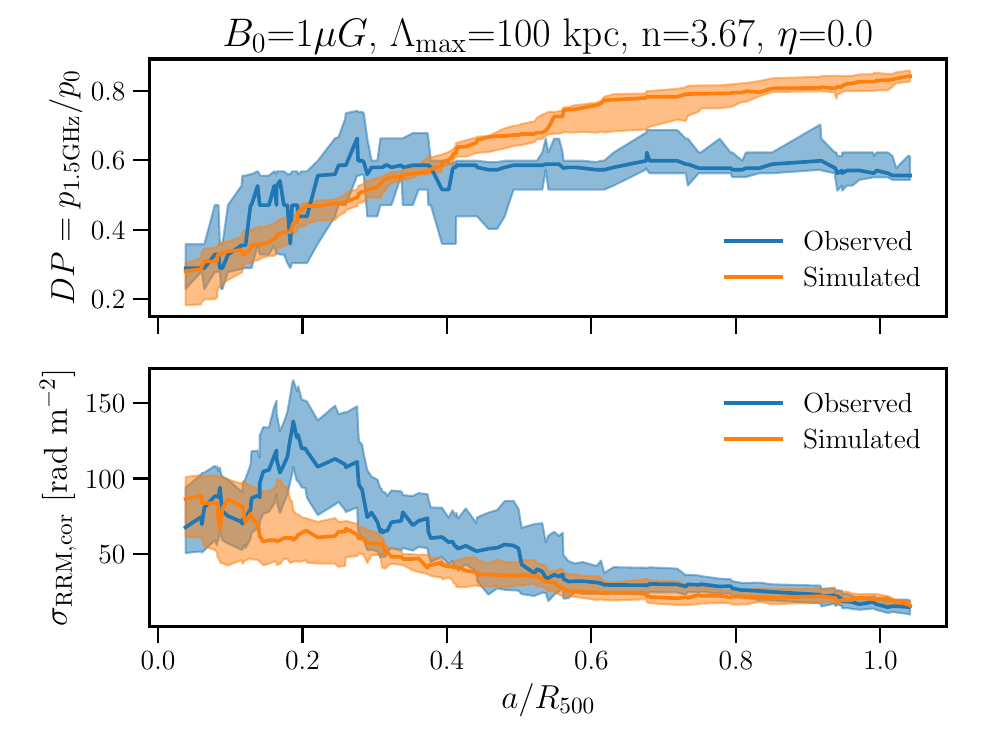}
    \caption{Model that minimizes $q_\mathrm{RM}$.} \label{fig:bestfitKolmogorov_eta00b}
  \end{subfigure}%
\caption{Comparison between observed and forward-modelled depolarisation and RM scatter for the model with $\eta=0.0$ that minimizes the q statistics as defined in Equation \ref{eq:q}.} \label{fig:bestfitKolmogorov_eta00}
\end{figure*}

\FloatBarrier

\section{Updated polarised source catalogue}\label{app:polcatalogue}



\begin{table*}[hbt]
\centering
\caption{First 30 rows of the catalogue of 819 polarised radio sources that were detected in this work.}
\label{tab:polarisedtable}
\resizebox{\textwidth}{!}{%
\begin{tabular}{@{}lllllllllllllllllllllll@{}}
\toprule
RA{[deg]} & DEC{[deg]} & Maj{[$^{\prime\prime}$]} & Min{[$^{\prime\prime}$]} & PA{[deg]} & $p_0$ & $\chi_0${[rad]} & RM{[rad m$^{-2}$]} & $\sigma_\mathrm{RM}${[rad m$^{-2}$]} & $I_0${[mJy]} & $\alpha$ & $\chi^2_\mathrm{QU}$ & $z_\mathrm{best}$ & $z_\mathrm{best}$ source & $\theta_p${[arcmin]} & $r/R_{500}$ & Cluster & RA$_\mathrm{opt}${[deg]} & DEC$_\mathrm{opt}${[deg]} & Multi component & Flagged & Note & RRM{[rad m$^{-2}$]} \\ \midrule
$19.55453\pm0.00009$ & $-27.16767\pm0.00010$ & $10.4\pm1.1$ & $6.11\pm0.35$ & $136\pm7$ & $0.13^{+0.04}_{-0.03}$ & $3.0^{+0.3}_{-0.3}$ & $20^{+6}_{-5}$ & $8^{+5}_{-3}$ & $5.0^{+0.2}_{-0.2}$ & $-0.5^{+0.1}_{-0.1}$ & 61 & $0.948 \pm 0.077$ & 2 & $0.19$ & $2.9$ & G212.97-84.04 & 19.5606 & -27.1622 & False & False &  & $7^{+7}_{-7}$ \\
$15.794198\pm0.000035$ & $-21.996243\pm0.000030$ & $7.03\pm0.31$ & $6.29\pm0.24$ & $117\pm16$ & $0.08^{+0.03}_{-0.02}$ & $1.0^{+0.3}_{-0.3}$ & $30^{+7}_{-6}$ & $13^{+3}_{-3}$ & $12.7^{+0.2}_{-0.2}$ & $-0.88^{+0.04}_{-0.04}$ & 52 & - & - & $0.16$ & $0.6$ & G149.55-84.16 & - & - & False & False & No optical counterpart. & $24^{+8}_{-7}$ \\
$41.4094860\pm0.0000022$ & $-20.6108492\pm0.0000024$ & $9.548\pm0.021$ & $8.841\pm0.018$ & $59.7\pm1.2$ & $0.104^{+0.001}_{-0.001}$ & $0.465^{+0.008}_{-0.008}$ & $13.6^{+0.1}_{-0.1}$ & $2.4^{+0.4}_{-0.4}$ & $149^{+2}_{-2}$ & $-0.68^{+0.03}_{-0.03}$ & 434 & - & - & $0.21$ & $3.5$ & G205.07-62.94 & - & - & False & True & No optical counterpart. & $8^{+3}_{-3}$ \\
$41.411690\pm0.000009$ & $-20.580840\pm0.000007$ & $9.18\pm0.08$ & $8.14\pm0.06$ & $101.3\pm2.9$ & $0.119^{+0.003}_{-0.003}$ & $1.64^{+0.03}_{-0.03}$ & $16.5^{+0.6}_{-0.6}$ & $1.6^{+1.0}_{-0.6}$ & $28.6^{+0.4}_{-0.4}$ & $-0.83^{+0.04}_{-0.04}$ & 144 & - & - & $0.20$ & $3.3$ & G205.07-62.94 & - & - & False & True & No optical counterpart. & $10^{+3}_{-3}$ \\
$40.48671\pm0.00010$ & $-28.63936\pm0.00021$ & $30.8\pm1.9$ & $10.4\pm0.4$ & $113.9\pm3.3$ & $0.28^{+0.02}_{-0.02}$ & $0.37^{+0.07}_{-0.07}$ & $8^{+1}_{-1}$ & $7.2^{+1.0}_{-0.9}$ & $16.1^{+0.5}_{-0.5}$ & $-1.61^{+0.07}_{-0.08}$ & 89 & $0.238 \pm 0.000$ & 1 & $0.11$ & $1.6$ & G222.97-65.69 & 40.4839 & -28.6379 & True & True & Radio Relic. & $3^{+3}_{-3}$ \\
$40.47708\pm0.00020$ & $-28.63432\pm0.00015$ & $20.7\pm2.0$ & $10.0\pm0.6$ & $124\pm6$ & $0.30^{+0.06}_{-0.05}$ & $0.2^{+0.2}_{-0.2}$ & $7^{+3}_{-3}$ & $10^{+2}_{-2}$ & $9.7^{+0.4}_{-0.4}$ & $-2.4^{+0.1}_{-0.1}$ & 80 & $0.238 \pm 0.000$ & 1 & $0.10$ & $1.5$ & G222.97-65.69 & 40.4839 & -28.6379 & True & True & Radio Relic. & $2^{+4}_{-4}$ \\
$40.48939\pm0.00025$ & $-28.65367\pm0.00018$ & $20.5\pm2.3$ & $14.8\pm1.4$ & $114\pm16$ & $0.14^{+0.01}_{-0.01}$ & $1.9^{+0.2}_{-0.2}$ & $6^{+3}_{-3}$ & $2.2^{+1.8}_{-0.8}$ & $9.6^{+0.4}_{-0.4}$ & $-1.1^{+0.1}_{-0.1}$ & 88 & $0.238 \pm 0.000$ & 1 & $0.11$ & $1.6$ & G222.97-65.69 & 40.4839 & -28.6379 & True & True & Radio Relic. & $1^{+4}_{-4}$ \\
$40.4898\pm0.0005$ & $-28.64934\pm0.00008$ & $46\pm4$ & $10.2\pm0.5$ & $96\pm4$ & $0.29^{+0.03}_{-0.03}$ & $1.7^{+0.1}_{-0.1}$ & $9^{+2}_{-2}$ & $6^{+1}_{-1}$ & $13.9^{+0.6}_{-0.6}$ & $-1.9^{+0.1}_{-0.1}$ & 131 & $0.238 \pm 0.000$ & 1 & $0.11$ & $1.6$ & G222.97-65.69 & 40.4839 & -28.6379 & True & True & Radio Relic. & $5^{+3}_{-4}$ \\
$40.47123\pm0.00030$ & $-28.63765\pm0.00029$ & $29.7\pm3.2$ & $17.2\pm1.5$ & $134\pm11$ & $0.19^{+0.04}_{-0.04}$ & $1.6^{+0.2}_{-0.2}$ & $7^{+3}_{-3}$ & $9^{+2}_{-2}$ & $20.3^{+0.7}_{-0.7}$ & $-2.5^{+0.1}_{-0.1}$ & 61 & $0.238 \pm 0.000$ & 1 & $0.09$ & $1.4$ & G222.97-65.69 & 40.4839 & -28.6379 & True & True & Radio Relic. & $2^{+4}_{-5}$ \\
$40.33575\pm0.00011$ & $-28.40285\pm0.00006$ & $10.4\pm1.0$ & $7.1\pm0.4$ & $104\pm9$ & $-$ & $-$ & $-$ & $-$ & $-$ & $-$ & - & $1.135 \pm 0.185$ & 2 & $0.25$ & $3.5$ & G222.97-65.69 & - & - & True & True & No optical counterpart. & $-$ \\
$40.33588\pm0.00004$ & $-28.41099\pm0.00004$ & $10.4\pm0.4$ & $7.16\pm0.18$ & $138\pm4$ & $0.034^{+0.005}_{-0.005}$ & $1.7^{+0.2}_{-0.2}$ & $17^{+3}_{-3}$ & $8^{+2}_{-1}$ & $52.3^{+0.7}_{-0.7}$ & $-0.96^{+0.03}_{-0.03}$ & 85 & $1.135 \pm 0.185$ & 2 & $0.24$ & $3.4$ & G222.97-65.69 & 40.3352 & -28.4072 & True & False &  & $13^{+4}_{-4}$ \\
$10.94092\pm0.00004$ & $-20.49858\pm0.00004$ & $9.5\pm0.4$ & $7.56\pm0.25$ & $143\pm8$ & $0.040^{+0.009}_{-0.006}$ & $0.6^{+0.1}_{-0.1}$ & $-10^{+3}_{-3}$ & $7^{+3}_{-2}$ & $17.1^{+0.3}_{-0.3}$ & $-0.71^{+0.04}_{-0.04}$ & 72 & $0.414 \pm 0.068$ & 2 & $0.14$ & $2.3$ & G106.73-83.22 & 10.9412 & -20.4988 & False & False &  & $-13^{+5}_{-5}$ \\
$10.79686\pm0.00014$ & $-20.53360\pm0.00005$ & $22.8\pm1.2$ & $11.9\pm0.4$ & $84\pm4$ & $0.21^{+0.02}_{-0.02}$ & $2.5^{+0.1}_{-0.1}$ & $-6^{+2}_{-2}$ & $7^{+1}_{-1}$ & $15.0^{+0.5}_{-0.5}$ & $-1.85^{+0.10}_{-0.10}$ & 89 & $0.281 \pm 0.013$ & 2 & $0.09$ & $1.4$ & G106.73-83.22 & 10.7947 & -20.5365 & False & False &  & $-9^{+5}_{-5}$ \\
$10.75281\pm0.00012$ & $-20.58744\pm0.00006$ & $18.1\pm1.2$ & $6.67\pm0.17$ & $155.7\pm2.4$ & $0.078^{+0.009}_{-0.008}$ & $2.4^{+0.1}_{-0.1}$ & $-11^{+2}_{-2}$ & $6^{+2}_{-1}$ & $18.3^{+0.4}_{-0.4}$ & $-1.12^{+0.05}_{-0.05}$ & 75 & $0.729 \pm 0.067$ & 2 & $0.09$ & $1.4$ & G106.73-83.22 & 10.7440 & -20.5799 & True & False &  & $-14^{+5}_{-5}$ \\
$10.74349\pm0.00005$ & $-20.62051\pm0.00012$ & $14.2\pm1.1$ & $6.54\pm0.22$ & $159.5\pm3.4$ & $0.0014^{+0.0003}_{-0.0002}$ & $0.2^{+0.2}_{-0.2}$ & $17^{+4}_{-4}$ & $8^{+3}_{-2}$ & $479^{+5}_{-5}$ & $-0.83^{+0.03}_{-0.02}$ & 184 & $0.281 \pm 0.272$ & 2 & $0.10$ & $1.5$ & G106.73-83.22 & 10.7432 & -20.6204 & False & True &  & $14^{+6}_{-6}$ \\
$10.743987\pm0.000028$ & $-20.57991\pm0.00004$ & $8.81\pm0.35$ & $7.17\pm0.22$ & $160\pm7$ & $0.0036^{+0.0008}_{-0.0006}$ & $2.0^{+0.2}_{-0.2}$ & $-157^{+4}_{-4}$ & $6^{+4}_{-2}$ & $134^{+2}_{-2}$ & $-0.53^{+0.03}_{-0.03}$ & 57 & $0.729 \pm 0.067$ & 2 & $0.10$ & $1.6$ & G106.73-83.22 & 10.7440 & -20.5799 & True & False &  & $-160^{+6}_{-6}$ \\
$10.73185\pm0.00011$ & $-20.57074\pm0.00017$ & $14.4\pm1.4$ & $11.5\pm1.0$ & $172\pm19$ & $0.09^{+0.02}_{-0.01}$ & $1.1^{+0.2}_{-0.2}$ & $-3^{+4}_{-3}$ & $6^{+3}_{-2}$ & $9.5^{+0.3}_{-0.3}$ & $-0.60^{+0.08}_{-0.08}$ & 73 & $0.729 \pm 0.067$ & 2 & $0.11$ & $1.8$ & G106.73-83.22 & 10.7440 & -20.5799 & True & False &  & $-6^{+5}_{-5}$ \\
10.8157 & -20.5492 & 8.2 & 8.2 & 0 & $0.19^{+0.05}_{-0.04}$ & $3.0^{+0.2}_{-0.2}$ & $-19^{+4}_{-4}$ & $6^{+3}_{-2}$ & $3.2^{+0.3}_{-0.3}$ & $-2.2^{+0.2}_{-0.2}$ & 71 & $0.603 \pm 0.041$ & 2 & $0.07$ & $1.0$ & G106.73-83.22 & 10.8167 & -20.5483 & False & False &  & $-22^{+6}_{-6}$ \\
$46.89273\pm0.00006$ & $-28.725980\pm0.000029$ & $7.1\pm0.5$ & $5.03\pm0.22$ & $106\pm7$ & $0.4^{+0.3}_{-0.3}$ & $0^{+1}_{-1}$ & $83^{+74}_{-75}$ & $41^{+5}_{-8}$ & $11.8^{+0.2}_{-0.2}$ & $-0.57^{+0.04}_{-0.04}$ & 120 & $0.270 \pm 0.020$ & 2 & $0.13$ & $1.9$ & G223.91-60.09 & 46.8925 & -28.7259 & False & False &  & $76^{+74}_{-75}$ \\
$46.83815\pm0.00017$ & $-28.68291\pm0.00004$ & $13.7\pm1.5$ & $6.18\pm0.30$ & $95\pm5$ & $0.03^{+0.02}_{-0.01}$ & $2.5^{+0.3}_{-0.3}$ & $40^{+9}_{-7}$ & $12^{+6}_{-3}$ & $20.1^{+0.3}_{-0.3}$ & $-0.85^{+0.04}_{-0.04}$ & 78 & $0.383 \pm 0.035$ & 2 & $0.07$ & $1.1$ & G223.91-60.09 & 46.8365 & -28.6812 & True & False &  & $33^{+10}_{-8}$ \\
$46.83576\pm0.00015$ & $-28.67959\pm0.00021$ & $17.2\pm2.2$ & $6.00\pm0.30$ & $145\pm5$ & $0.07^{+0.05}_{-0.02}$ & $2.5^{+0.5}_{-0.5}$ & $43^{+19}_{-16}$ & $24^{+7}_{-4}$ & $25.1^{+0.4}_{-0.3}$ & $-0.76^{+0.03}_{-0.03}$ & 71 & $0.383 \pm 0.035$ & 2 & $0.06$ & $1.1$ & G223.91-60.09 & 46.8365 & -28.6812 & True & False &  & $35^{+20}_{-16}$ \\
$46.83814\pm0.00005$ & $-28.68534\pm0.00005$ & $7.2\pm0.5$ & $5.38\pm0.26$ & $138\pm9$ & $0.04^{+0.51}_{-0.02}$ & $2.1^{+0.6}_{-0.6}$ & $79^{+18}_{-48}$ & $17^{+2283}_{-5}$ & $11.9^{+0.2}_{-0.2}$ & $-0.71^{+0.04}_{-0.04}$ & 54 & $0.383 \pm 0.035$ & 2 & $0.07$ & $1.1$ & G223.91-60.09 & 46.8365 & -28.6812 & True & True &  & $72^{+19}_{-48}$ \\
$357.900782\pm0.000032$ & $-25.938221\pm0.000018$ & $7.68\pm0.28$ & $5.76\pm0.15$ & $102\pm5$ & $0.038^{+0.010}_{-0.007}$ & $0.4^{+0.2}_{-0.2}$ & $-26^{+4}_{-4}$ & $7^{+3}_{-2}$ & $16.5^{+0.2}_{-0.2}$ & $-0.55^{+0.04}_{-0.04}$ & 63 & $0.187 \pm 0.133$ & 3 & $0.14$ & $2.0$ & G034.03-76.59 & 357.9006 & -25.9388 & False & False &  & $-34^{+6}_{-6}$ \\
$156.14934\pm0.00006$ & $-27.20391\pm0.00004$ & $7.9\pm0.6$ & $6.10\pm0.31$ & $112\pm10$ & $0.04^{+0.02}_{-0.01}$ & $1.7^{+0.4}_{-0.4}$ & $48^{+12}_{-10}$ & $16^{+5}_{-3}$ & $13.3^{+0.2}_{-0.2}$ & $-0.39^{+0.04}_{-0.04}$ & 76 & $0.258 \pm 0.163$ & 3 & $0.17$ & $2.4$ & G266.84+25.07 & 156.1492 & -27.2033 & False & False &  & $49^{+21}_{-20}$ \\
$156.02742\pm0.00005$ & $-27.331459\pm0.000032$ & $8.4\pm0.4$ & $7.05\pm0.27$ & $90\pm10$ & $0.06^{+0.01}_{-0.01}$ & $0.4^{+0.2}_{-0.2}$ & $-127^{+6}_{-6}$ & $15^{+3}_{-2}$ & $10.8^{+0.2}_{-0.2}$ & $-0.78^{+0.04}_{-0.04}$ & 95 & - & - & $0.08$ & $1.1$ & G266.84+25.07 & - & - & False & False & No optical counterpart. & $-124^{+21}_{-21}$ \\
$155.935295\pm0.000027$ & $-27.34246\pm0.00004$ & $8.82\pm0.34$ & $7.21\pm0.22$ & $11\pm7$ & $0.07^{+0.01}_{-0.01}$ & $1.4^{+0.2}_{-0.2}$ & $-132^{+7}_{-7}$ & $20^{+2}_{-2}$ & $17.3^{+0.3}_{-0.3}$ & $-0.55^{+0.03}_{-0.03}$ & 113 & - & - & $0.08$ & $1.2$ & G266.84+25.07 & - & - & False & False & No optical counterpart. & $-129^{+19}_{-19}$ \\
$155.723060\pm0.000033$ & $-27.180613\pm0.000032$ & $9.01\pm0.29$ & $8.66\pm0.26$ & $116\pm33$ & $0.025^{+0.004}_{-0.003}$ & $0.9^{+0.1}_{-0.1}$ & $-9^{+3}_{-2}$ & $5^{+2}_{-2}$ & $43.3^{+0.6}_{-0.6}$ & $-1.01^{+0.04}_{-0.04}$ & 138 & - & - & $0.24$ & $3.4$ & G266.84+25.07 & - & - & False & False & No optical counterpart. & $-11^{+25}_{-25}$ \\
$172.93573\pm0.00008$ & $-19.87149\pm0.00008$ & $8.0\pm0.7$ & $7.7\pm0.6$ & $(1.4\pm0.8)\times10^2$ & $0.3^{+0.6}_{-0.2}$ & $0^{+1}_{-1}$ & $-142^{+958}_{-739}$ & $43^{+192880647}_{-17}$ & $2.6^{+0.2}_{-0.2}$ & $-0.7^{+0.1}_{-0.1}$ & 80 & $0.291 \pm 0.025$ & 3 & $0.09$ & $1.3$ & G278.60+39.17 & 172.9325 & -19.8793 & True & True &  & $-132^{+958}_{-739}$ \\
$172.932507\pm0.000035$ & $-19.87992\pm0.00007$ & $13.5\pm0.6$ & $8.47\pm0.23$ & $20\pm4$ & $0.044^{+0.005}_{-0.005}$ & $2.7^{+0.2}_{-0.2}$ & $-0^{+3}_{-3}$ & $4^{+2}_{-1}$ & $10.0^{+0.3}_{-0.3}$ & $-0.95^{+0.06}_{-0.06}$ & 152 & $0.291 \pm 0.025$ & 3 & $0.08$ & $1.2$ & G278.60+39.17 & 172.9325 & -19.8793 & True & True &  & $10^{+11}_{-11}$ \\
$172.93171\pm0.00005$ & $-19.88390\pm0.00005$ & $9.1\pm0.5$ & $8.00\pm0.35$ & $139\pm16$ & $0.08^{+0.02}_{-0.02}$ & $2.8^{+0.3}_{-0.3}$ & $-115^{+7}_{-7}$ & $14^{+3}_{-2}$ & $4.8^{+0.2}_{-0.2}$ & $-0.69^{+0.08}_{-0.08}$ & 141 & $0.291 \pm 0.025$ & 3 & $0.08$ & $1.2$ & G278.60+39.17 & 172.9325 & -19.8793 & True & False &  & $-105^{+13}_{-13}$ \\ \bottomrule
\end{tabular}
}
\tablefoot{The columns are identical to \citet{Osinga2022}, except the newly added column RRM, which is the observed RM corrected for the Galactic value. If uncertainties are not available, they are not given. The full table is available in electronic form at the CDS.}
\end{table*}

\end{appendix}

\end{document}